\documentclass[prd,showpacs,nofootinbib]{revtex4-2}
\usepackage{amssymb, amsmath,amsthm,graphicx,hyperref,subfigure}

\begin{document}
\title{1D Supergravity FLRW Model of Starobinsky}
\author{N.E. Mart\'inez-Pérez}
\email{nephtalieliceo@outlook.com}
	
\author{C. Ramírez}
\email{cramirez@fcfm.buap.mx}
\affiliation{Benem\'erita Universidad Aut\'onoma de Puebla, Facultad de Ciencias
F\'{\i}sico Matem\'aticas, P.O. Box 165, 72000 Puebla, M\'exico.}

\author{V. Vázquez-Báez}
\email{manuel.vazquez@correo.buap.mx}
\affiliation{Benemérita Universidad Autónoma de Puebla, Facultad de Ingeniería, 72000 Puebla, México.}
\date{\today}
	
\begin{abstract}
We study two homogeneous supersymmetric extensions for the $f(R)$ modified gravity model of Starobinsky with the FLRW metric. The actions are defined in terms of a superfield $\mathcal{R}$ that contains the FLRW scalar curvature. One model has N=1 local supersymmetry, and its bosonic sector is the Starobinsky action; the other action has N=2, its bosonic sector contains, in additional to Starobinsky, a massive scalar field without self-interaction. As expected, the bosonic sectors of these models are consistent with cosmic inflation, as we show by solving numerically the classical dynamics. Inflation is driven by the $R^2$ term during the large curvature regime. In the N=2 case, the additional scalar field remains in a low energy state during inflation. Further, by means of an additional superfield, we write equivalent tensor-scalar-like actions from which we can give the Hamiltonian formulation.
\end{abstract}
\maketitle

\section{Introduction}
The model of Starobinsky is formulated as a string-inspired effective action, with~higher-order curvature terms arising from corrections due to quantized conformally covariant matter fields \citep{starobinsky1980,starobinsky1983}. The~relevant action for FLRW cosmology is given by the simplest version of the model, with~only quadratic scalar curvature \citep{kaneda},  $L_S=\sqrt{-g}(R+\frac{\alpha}{6}R^2)$. The~$R^2$ term is able to drive inflation in the large curvature regime $R\gg M^2=\alpha^{-1}$. Nowadays, it continues to be a viable inflationary model. The~predicted spectral index and tensor-to-scalar ratio, $n_s\approx 0.96$, $r\approx 0.004$, respectively, with~$M$ of the order of $10^{13}$ GeV, are in good agreement with current observational data \citep{mukhanov,defelice2010,ketovn}.

The model with quadratic scalar curvature is also the simplest of the $f(R)$ theories of modified gravity. These theories have drawn a lot of attention since they might account for, for example,~inflation as~well as the late-time accelerated cosmic expansion \citep{nojiri,defelice2010, sotiriou}. This is due to a higher derivative scalar degree of freedom, so-called scalaron, which is, therefore, of~a purely gravitational nature, at~least from a macroscopic point of view \citep{alexandre}. Although~the Starobinsky theory is higher derivative, it has the relevant feature that it clearly does not have ghosts and is stable \citep{woodard,chen}. This property can be seen from its relation through a Weyl rescaling, with~Einstein's theory with a minimally coupled scalar model  \citep{whitt, cecotti}. The~potential of the scalar in this model is consistent with large-field~inflation. 

Supersymmetric theories \citep{wessbagger} have been the subject of intense study for many years, creating a wide expectation for present LHC energies. As~it is well known, it could not be confirmed. Nevertheless, it continues to be considered, in~particular in view of recent results \citep{lhc}. On~the other side, there is no reason why supersymmetry should break at LHC scales; it can be much higher \citep{ketovn}. Inflation can be considered from Planck energy scale \citep{planck}. In~fact, the~physically relevant inflation, taking place after horizon exit of the observable universe, is typically several orders of magnitude below $M_P$ \citep{lyth}, but~still sufficiently high to be considered in a context of supergravity, as~a possible effective theory, or~even as an ultraviolet completion of quantum gravity \citep{ellis1982,ellis2013}.

Regarding the model of Starobinsky as an $f(R)$ action, one might expect $f(\mathcal{R})$, with~$\mathcal{R}$ the four-dimensional chiral curvature superfield, to~provide an adequate supersymmetrization \citep{ketov2013-1,diamandis}. The~connection between $f(\mathcal{R})$ and $f(R)$ is, however, not straightforward, due to auxiliary fields satisfying the rather involved algebraic equations of motion \citep{ketov2011}. More general actions depending not only on $\mathcal{R}$, but~its supersymmetric covariant derivatives, have been used to properly embed the model of Starobinsky into N=1, 2 4D supergravity~\citep{ketov2013-2,ketov2014,terada,ketovn}. 

On the side of scalar--tensor theories, the~main obstacle for the embedding of inflation into supergravity is finding a suitable scalar potential (it has to be sufficiently flat, at~least in a certain field direction in multi-field models) that generates the proper amount of inflation in a consistent way with the observations as~well as with the predictions of the standard model of cosmology on the early universe \citep{mcallister,stewart}. The~generic scalar potential in N=1 4D supergravity is too steep for slow-roll inflation \citep{ketov2011}. Some methods have been put forward to, for instance, make the potential independent of some field such that it is flat along its direction \citep{kawasaki, kallosh2011pr}. The~behavior of fermionic fields is also relevant for precise predictions; in~this regard, a~new kind of inflationary models with a simplified fermionic sector, i.e.,~with no inflation, is considered by~means of nilpotent superfields \citep{carrasco2015,terada2021}. Flat directions of the scalar potential can also be achieved with non-minimal coupling as in NMSSM models (Higgs inflation) \citep{kallosh2010, kallosh2011}. With~respect to the specific scalar potential of Starobinsky in its dual scalar-tensor form, its supersymmetrization is addressed in super-conformal theory \citep{kallosh2013} and no-scale supergravity \citep{ellis2013, tamvakis}.

As supersymmetry involves fermions, its study requires quantum theory, unless~only the scalar potential is considered as~is usually done in supergravity and superstring cosmology \citep{stewart}. On~the other side, the~homogeneity at the beginning of inflation has led to the consideration of homogeneous supergravity formulations \citep{obregon0}; see also \citep{moniz}. Due to the dimensional reduction to one dimension, these formulations lack Lorentz constraints, which were implemented by hand in \citep{qc-sugra2,obregon1}; otherwise the theory has a quite complicated constraint algebra \citep{damour}. In~view of that, in~\citep{tkach} a supersymmetric extension of the FLRW model was proposed, considering, as~usual, the~rescaled scale factor with length dimension. As~well as in 4D supergravity, the~basic superfield in the Lagrangian does not follow from the geometric considerations, but~is ad hoc \citep{wessbagger,tkach}.

In this work, we construct supersymmetric generalizations for the FLRW model of Starobinsky in its modified gravity form, using the `new' superspace formulation \citep{wessbagger,ramirez} for homogeneous supergravity \citep{garcia}, i.e.,~depending only on time.  As~FLRW involves gauge fixing, the~supersymmetric extension must keep it. This formulation is fully geometric, and~can be seen as a minimal dimensional reduction from 4D that keeps only the minimum elements for supersymmetry. In~fact, for~a dimensional reduction to one dimension, to~each fermionic component of the supersymmetric charge corresponds, in one dimension, one supersymmetry. In~this view, we consider N=1 and N=2 one-dimensional supergravity (as a complex representation, N=2 can be taken as N=1). Thus, an~extension back to spatial dimensions is straightforward, for~instance, to include perturbations.  Actions based on four-dimensional supergravity and more fundamental theories may contain a vast number of additional both dynamical and auxiliary fields, which usually renders the sole identification of the scalar potential a nontrivial task~\cite{mcallister}. In~our case, the~small number of degrees of freedom involved allows us to write both bosonic and fermionic sectors of the Lagrangians and Hamiltonians in full detail, not just the leading terms of certain Taylor~expansion. 

In Section~\ref{s2}, we discuss the effective 1D FLRW model of Starobinsky. In~order to substantiate the following sections, we discuss the Ostrogradsky formulation, and~show the canonical transformations that relate its Hamiltonian to~the two Hamiltonians of the scalar--tensor formulations usually related to the Starobinsky model. One of these scalar--tensor formulations is the one in standard form, which is manifestly stable, and~has a large-field inflationary potential. The~other scalar--tensor form, of~the BF-type, is the one we extend by supersymmetry. In~Section~\ref{n1}, we present a formulation based on the simplest possible superspace compatible with time-dependent supersymmetry transformations \citep{ramirez2008}. It has local coordinates $(t, \Theta)$, where $t$ is the time coordinate and $\Theta$ is a real Grassmann number \citep{henneaux}. Superfields are very simple: they only contain one real boson scalar and one real fermion scalar, and~there are no auxiliary fields. Despite the simplicity of this superspace, we can write a supersymmetric Lagrangian whose bosonic part contains exactly $R+\frac{\alpha}{6}R^2$. In~this construction, the Grassmann parity of the curvature superfield is odd. In~Section~\ref{n2}, we make use of a complex superspace having local coordinates $(t, \Theta, \bar{\Theta})$. In~this case, superfields contain twice as many components: two real scalar bosons and one complex scalar fermion. Usually, one of the bosons is an auxiliary supersymmetric field that is used to ensure the off-shell closure of the supersymmetry algebra \citep{belluci,wessbagger}. As~we will show, in~higher-derivative models it becomes a dynamical field on its own. In~this case, the curvature superfield is of even parity and the action can, in~principle, be generalized to the superspace of 4D supergravity. We propose an action depending on the curvature superfield and its supersymmetric covariant derivatives in a similar fashion to the proper embeddings of Starobinsky into 4D supergravity \citep{ketov2013-2,terada,cecotti}. By~choosing a specific superpotential, we obtain a model whose bosonic part contains Starobinsky and a massive scalar field. We present numerical solutions to the pure bosonic equations of motion, which exhibit inflation driven by the quadratic curvature term, whereas the scalar field is pushed to the minimum energy state (see \citep{ketovn} for another example of a two-field model, where the effect on inflation of both fields is detailed). In~Section~\ref{cf}, we present the Hamiltonian formulation, including bosons and fermions, of~the equivalent scalar--`tensor' actions for the two models constructed. Unlike ordinary supersymmetric theories, these actions contain quadratic terms in the fermionic velocities; nonetheless, the~ordinary Hamiltonian formulation can be carried out as normal. Section~\ref{concl} is dedicated to the conclusions and final remarks. In~Appendix \ref{app}, we summarize the basic results about the superspace formalism following references \citep{ramirez,garcia, wessbagger}, and~in Appendix \ref{app2}, we provide some lengthy~equations. 

\section{FLRW Model of~Starobinsky}\label{s2}
The FLRW geometry corresponds to a spacetime with spatial 3-surfaces of uniform curvature $^3R=6ka^{-2}$, where $k$ takes the values $-$1, 0 or 1. The~line element, in~comoving coordinates ($t,r,\theta,\phi$), reads $ds^2=-N^2(t)dt^2+a^2(t)\left((1-kr^2)^{-1}dr^2+r^2 d\Omega_2^2\right)$. The~scale factor $a(t)$ is the only degree of freedom. Actually, it is the Hubble factor $H=\frac{\dot{a}}{a}$, which is directly observable. The~field equations for a spatially homogeneous and isotropic $f(R)$ universe, which generalize the usual Friedmann and acceleration (Raychaudhuri) equations, can be directly obtained from the $f(R)$ action evaluated at the FLRW geometry, after~integration of the spatial coordinates. With~the FLRW metric, the~scalar curvature $R$~becomes the following:
\begin{equation}\label{curvature}
R=6 \left(-\frac{\dot{N} \dot{a}}{N^3 a}+\frac{\ddot{a}}{N^2 a}+\frac{\dot{a}^2}{N^2 a^2}+\frac{k}{a^2}\right).
\end{equation}

Substituting these expressions into $\frac{1}{2\kappa^2} \int d^4 x \sqrt{-g} R$, and~integrating the spatial coordinates over a suitable region, yields the FLRW action. 
In this work, we are mainly interested in the model of Starobinsky $L_S=(2\kappa^2)^{-1} Na^3 [R+(\alpha/6) R^2]$,  with~$k=0$, that is, the following:
\begin{equation}\label{starobinsky}
L_S=\frac{3}{\kappa^2} N a^3 \left[-\frac{\dot{N} \dot{a}}{N^3 a}+\frac{\ddot{a}}{N^2 a}+\frac{\dot{a}^2}{N^2 a^2}+\alpha \left(-\frac{\dot{N} \dot{a}}{N^3 a}+\frac{\ddot{a}}{N^2 a}+\frac{\dot{a}^2}{N^2 a^2}\right)^2\right].
\end{equation}

This higher derivative action is our starting point for the supergravity version; hence, it is important to verify its consistency. In~order to do it, we derive the Ostrogradsky--Hamiltonian formulation, and~show that it is related through canonical transformations to two well-known scalar--tensor formulations: one of the BF type, with~the scalar playing the role of the scalar curvature, and~the other the standard Hamiltonian obtained by means of a Weyl~transformation.

\subsection{Ostrogradsky--Hamiltonian~Formulation}

The canonical formulation of the FLRW model of Starobinsky can be obtained directly from the higher derivative action (\ref{starobinsky}) following the method of Ostrogradsky (see, for example,~\cite{woodard} for a review). For~simplicity, we integrate by parts the second derivative on the linear curvature term in (\ref{starobinsky}). In~this formalism, there are eight canonical variables, namely the following:
\begin{subequations}
	\begin{eqnarray}
	&&a, \\
	&&p_a \equiv \frac{\partial L_S}{\partial \dot{a}}-\frac{d}{dt} \frac{\partial L_S}{\partial \ddot{a}}=-\frac{6}{\kappa^2} \left[\frac{a \dot{a}}{N}+\frac{\alpha a^2}{N} \frac{d}{dt} \left(\frac{\ddot{a}}{N^2 a}-\frac{\dot{N} \dot{a}}{N^3 a}+\frac{\dot{a}^2}{N^2 a^2}\right)\right], \\
	&&A\equiv \dot{a},\\
	&&p_A\equiv \frac{\partial L_S}{\partial \ddot{a}}=\frac{6\alpha}{\kappa^2} \frac{a^2}{N} \left(\frac{\ddot{a}}{N^2a}-\frac{\dot{N} \dot{a}}{N^3 a}+\frac{\dot{a}^{2}}{N^2 a^2}\right),\\
	&&N,\\
	&&p_N\equiv \frac{\partial L_S}{\partial \dot{N}}-\frac{d}{dt} \frac{\partial L_S}{\partial \ddot{N}}=-\frac{6\alpha}{\kappa^2} \frac{a^2 \dot{a}}{N^2} \left(\frac{\ddot{a}}{N^2a}-\frac{\dot{N} \dot{a}}{N^3 a}+\frac{\dot{a}^{2}}{N^2 a^2}\right), \label{pNos} \\
	&&n \equiv \dot{N}, \\
	&&p_n \equiv \frac{\partial L_S}{\partial \ddot{N}}=0. \label{pn}
	\end{eqnarray}
\end{subequations}

Since (\ref{starobinsky}) does not depend on $\ddot{N}$, we obtain a primary constraint (\ref{pn}). It can already be seen that $p_N$ can be written in terms of $A$ and $P_A$, which would yield another~constraint.

The Hamiltonian of Ostrogradsky is defined by the Legendre~transformation as follows:
\begin{subequations}
	\begin{eqnarray}
	&&H_{\text{Ost}} \equiv A p_a+\dot{A} p_A+n p_N+\dot{n} p_n-L_S \label{ostro1}\\
	&&\ \ \ \ \ \ \ \ \ =A p_a+\frac{3}{\kappa^2} \frac{a A^2}{N}+\frac{\kappa^2}{12 \alpha} \frac{N^{3} p_A^2}{a}-\frac{A^{2} p_A}{a}+n\left(\frac{A p_A}{N}+p_N\right)+\dot{n} p_n. \label{ostro2}
	\end{eqnarray}
\end{subequations}

At first sight, $H_{\text{Ost}}$ is not bounded due to terms being linear in the momenta $p_a$ and $p_N$ \mbox{(\ref{ostro1})}. However, the~model of Starobinsky is known to be stable. This is due to the system (\ref{starobinsky}) being degenerate in the sense that the matrix $(\partial^2 L_S/\partial \ddot{q}_i \partial \ddot{q}_j)$ is not invertible. Otherwise, it would have fourth-order equations of motion for both $a$ and $N$, but~only $a$ obtains a higher-derivative~equation.

As we obtain a primary constraint, we proceed following Dirac's standard procedure (see~\cite{henneaux}). The~result is that there are three first-class~constraints as follows:
\begin{subequations}
	\begin{eqnarray}
	&& 0 \approx C \equiv  p_n \label{c01}, \\
	&& 0\approx D\equiv \frac{A p_A}{N}+p_N, \label{c1}\\
	&& 0\approx H_0\equiv \frac{A p_a}{N}+\frac{3}{\kappa^2} \frac{a A^2}{N^2}+\frac{\kappa^2}{12 \alpha} \frac{N^{2} p_A^2}{a}-\frac{A^{2} p_A}{N a} \label{h0ostr}.
	\end{eqnarray}
\end{subequations}
(\ref{c1}) follows from the conservation in time of (\ref{c01}), whereas (\ref{h0ostr}) follows from the conservation of (\ref{c1}). The~time derivative of (\ref{h0ostr}) vanishes identically and no more constraints arise. Therefore, the~total Hamiltonian vanishes as a constraint $0 \approx H=NH_0+n D+\nu C$, where $N$, $n$ and $\nu$ remain arbitrary. In~order to understand the Hamiltonian constraint $H_0$, taking into account that $p_A$ is proportional to $R$, we make a canonical transformation $p_\phi=-A$, and~$\phi=-p_A$. Thus, the following holds:
\begin{equation}
H_0=\frac{1}{N}\left(\frac{3a}{\kappa^2 N} +\frac{\phi}{a}\right) p_\phi^2-\frac{1}{N} p_\phi p_a+\frac{\kappa^2}{12 \alpha} \frac{N^2\phi^2}{a},
\end{equation}
which requires further analysis to see if it contains ghosts. As~we show in the following, there is a further canonical transformation that puts it in a more familiar form. The~other two constraints cannot affect~stability.

\subsection{Scalar-Tensor~Formulation}\label{bef}

There are other well-known ways to reduce the order of a Lagrangian by means of additional fields that keep track of the higher derivative terms, for~instance, by using Lagrange multipliers~\cite{vilenkin}, or~by actions of the BF type as follows:
\begin{equation}\label{firststar}
L_S^\phi =\frac{N {a}^{3}}{2\kappa^2} \left[(1+2 \alpha \phi) R-6\alpha {\phi}^{2} \right].
\end{equation}

Further, substituting $R$, and~integrating by parts yields the following:
\begin{eqnarray}\label{lagphi}
L_S^\phi=\frac{3}{\kappa^2} Na^3 \left[-\frac{\dot{a}^{2}}{a^2 N^2} \left(1+2\alpha \phi \right)-2\alpha \frac{\dot{a}\dot{\phi}}{aN^2}-\alpha {\phi}^{2}\right] .
\end{eqnarray}

Its Hamiltonian is $H^\phi=N H_0+\mu p_N$, where the Hamiltonian constraint is the following:
\begin{equation}\label{hamilphi}
H_0^\phi=-\frac{\kappa^2}{6\alpha} \frac{p_{a} p_\phi}{a^2}+\frac{\kappa^2}{12 \alpha^2} (1+2\alpha \phi) \frac{p_\phi^{2}}{a^3}+\frac{3\alpha}{\kappa^2} {a}^{3} {\phi}^{2}.
\end{equation}

As can be expected, this approach and that of Ostrogradsky give equivalent Hamiltonian constraints (\ref{h0ostr}) and (\ref{hamilphi}), as~they are related by the canonical~transformation as follows:
\begin{subequations}\label{canonical1}
\begin{align}
&a=a, && p_a^{\text{Ost}}=p_a-2\frac{\phi p_\phi}{a}, \\
&A=-\frac{\kappa^2}{6 \alpha} \frac{N p_\phi}{a^2}, && p_A=-\frac{6\alpha}{\kappa^2} \frac{a^2}{N} \phi, \\
&N=N, && p^{\text{Ost}}_N=p_N+\frac{\phi p_\phi}{N}.
\end{align}
\end{subequations}

In fact, also the other constraints transform into each other by this transformation, namely $p_N=N^{-1} A p_A+p_N^{\text{Ost}}=D$. Here, $ p_a^{\text{Ost}}$ and $p_N^{\text{Ost}}$ denote the corresponding momenta in the Ostrogradsky~formalism.

Both versions of the Hamiltonian constraint, (\ref{hamilphi}) or (\ref{h0ostr}), yield the same nontrivial relation that, expressed in configuration space variables, can be recognized as the generalized Friedmann equation for the model of Starobinsky, which in the flat case $k=0$ can be written as a second-order differential equation for the Hubble factor. In~the gauge $N=1$, it~reads as follows:
\begin{equation}
\ddot{H}-\frac{\dot{H}^{2}}{2H}+3H \dot{H}+\frac{1}{2} M^2 H=0 \label{friedmannstarobinsky}.
\end{equation}	

Stable inflationary dynamics, for~which the model of Starobinsky is greatly appreciated in cosmology, can be obtained from (\ref{friedmannstarobinsky}) \cite{defelice2010}. See also~\cite{ketovn} for a recent review of Starobinsky~inflation.

\subsection{Standard~Formulation}

It is well known that by means of a Weyl rescaling of the gravitational field plus a redefinition of the scalaron field, say $\phi$, any $f(R)$ action can be written as standard Einstein gravity with a scalar minimally coupled, with~a large field inflationary potential; see e.g., \cite{nojiri}. It is interesting to note that the corresponding canonical formulation in the FLRW case can be obtained from (\ref{hamilphi}) by the canonical~transformation as follows: \vspace{6pt}
\begin{subequations}\label{conformal}
\begin{align}
&\phi=(2 \alpha)^{-1} (e^{c \varphi}-1), && p_\phi=\alpha e^{-c \varphi} (\tilde{a} p_{\tilde{a}}+\tilde{N} p_{\tilde{N}})+\frac{2\alpha}{c} e^{- c \varphi} p_\varphi, \\
&a=e^{-\frac{1}{2} c\varphi} \tilde{a}, &&p_a=e^{\frac{c}{2} \varphi} p_{\tilde{a}}, \\
&N=e^{-\frac{1}{2} c\varphi} \tilde{N}, && p_N=e^{\frac{c}{2} \varphi} p_{\tilde{N}}
\end{align}
\end{subequations}

\noindent with $c^2=\frac{2}{3}\kappa^2$. Indeed, applying (\ref{conformal}) transforms (\ref{hamilphi}) into the following:
\begin{equation}\label{frwstar}
0\approx H_0^\varphi=-\frac{{\kappa}^{2}}{12} \frac{p_{\tilde{a}}^{2}}{\tilde{a}}+\frac{p_\varphi^{2}}{2 \tilde{a}^3}+\frac{3 M^2}{4 \kappa^2} \tilde{a}^3 (1-e^{-c \varphi})^2.
\end{equation}

In lagrangian terms, (\ref{frwstar}) corresponds to an ordinary Friedmann equation, which is complemented by the second-order equation of motion for $\varphi$.

Therefore, by~combining the transformations (\ref{canonical1}) and (\ref{conformal}), one can pass directly from the Ostrogradsky Hamiltonian (\ref{h0ostr}) to (\ref{frwstar}).

Now, the~Weyl rescaling on which (\ref{conformal}) is based takes us to the Einstein frame~\cite{nojiri}. We regard the frame of Starobinsky's modified gravity (generically called Jordan) as physical, such that inflation has a gravitational origin. Additionally, since $c \varphi=\ln (1+2\alpha \phi)$, the~transformation (\ref{conformal}) becomes singular at a sufficiently negative curvature (recalling $\phi=R/6$). Thus, for~example, it might be convenient to perform quantization of the supersymmetric models constructed in the following sections (see Section~\ref{cf}), while staying in the Jordan frame~\cite{hawkingluttrell}. 

Finally, although~the Hamiltonian constraint (\ref{hamilphi}) is already suitable for quantization~\cite{vilenkin}, one might prefer to have it in canonical form. This can be done by the canonical~transformation as follows:
\begin{subequations}
	\begin{align}
	&a=b+\varphi, && p_a=\frac{b p_b+\varphi p_\varphi}{b+\varphi}, \\
	&\phi=\frac{-1}{\alpha} \frac{\varphi}{b + \varphi}, && p_\phi=\alpha (b+\varphi) (p_b-p_\varphi),
	\end{align}	
\end{subequations}

\noindent which diagonalizes the kinetic part of (\ref{hamilphi}) such that it reads as follows:
\begin{equation}
H_0^\phi= \frac{\kappa^2}{12} (-p_b^2+p_\varphi^2)+\frac{3}{\kappa^2 \alpha} \varphi^2 (b+\varphi)^2.
\end{equation}

\section{N=1 Locally Supersymmetric~Action}\label{n1}
In this section, we construct the first example of a supersymmetric extension of the FLRW model of Starobinsky (\ref{starobinsky}). For~this, we use the simplest possible superspace compatible with time-dependent supersymmetry transformations~\cite{ramirez2008}. Its basic features are summarized in Appendix A; we recall here that it has local coordinates ($t, \Theta$), where $\Theta$ is a real odd parity Grassmann number. Superfields have only two components (\ref{realsuper}): one real boson scalar and one real fermion~scalar.

We define the real scale factor superfield (complex conjugation reverses the order, therefore, the product of two real odd parity Grassmann numbers is imaginary) as follows:
\begin{equation}\label{A}
\mathcal{A}(t,\Theta)=a(t) [1+i\Theta \lambda(t)].
\end{equation}
where, for~convenience, the~component expansion differs from the standard one \mbox{(\ref{realsuper})}. However, the~superfield transformation of (\ref{A}) is the usual one (see~\citep{garcia}), and~the transformation of its components can be obtained from (\ref{realtrans}). We~obtain the following:
\begin{align}\label{realscalet}
\delta_\zeta a=-i \zeta a \lambda, &&	\delta_\zeta \lambda=\zeta\left(\frac{\dot{a}}{aN}-i \psi \lambda\right). 
\end{align}

We define the $k=0$ curvature superfield (non-vanishing spatial curvature can be introduced via an interaction with a Golstino field $\beta(t)$. Specifically, we add to the lagrangian density in (\ref{linear}) a term $-k\mathcal{EA} \mathcal{B}$, where $\mathcal{B}=\beta+\Theta \left(-1+i\beta \psi+i \beta N^{-1} \dot \beta\right)$ is a Goldstino superfield~\cite{ramirez2008}) as follows:
\begin{equation}\label{superreal}
\mathcal R=i\mathcal{A}^{-1}\nabla_\tau \nabla_\theta \mathcal{A}+i \mathcal{A}^{-2} \nabla_\tau \mathcal{A} \nabla_\theta \mathcal{A},
\end{equation}
where covariant derivatives are given in (\ref{realcov}). Thus, we have the following:
\begin{equation}\label{superreal2}
	\mathcal{R}=-\frac{2\dot{a}}{Na} \lambda-\frac{\dot{\lambda}}{N}-\frac{\dot{a}}{Na} \psi+\Theta \left(-\frac{\dot{N}\dot{a}}{N^3a}+\frac{\ddot{a}}{N^2 a}+\frac{\dot{a}^2}{N^2 a^2}+i \frac{\lambda \dot{\psi}}{N}-\frac{6i \dot{a}}{Na} \psi \lambda-2i \frac{\psi \dot{\lambda}}{N}+2i \frac{\lambda \dot{\lambda}}{N}\right)
\end{equation}
and for the ordinary (linear) action we have the following:
\begin{equation}\label{linear}
L_1=\frac{3}{\kappa^2} \int d\Theta \mathcal{E} \mathcal{A}^3 \mathcal{R} \doteq \frac{3{a}^{3}}{\kappa^2} \left(\frac{\ddot{a}}{a}+\frac{\dot{a}^2}{a^2}-i \lambda \dot{\lambda}\right),
\end{equation}
where $\mathcal{E}$ is the density superfield given in (\ref{realdensity}). 
Since $\mathcal{R}$ has odd parity ($\mathcal{R}^2=0$), we cannot construct higher-order polynomials of it. However, it is possible to construct an exact supersymmetric model of Starobinsky by considering the product of $\mathcal{R}$ and its supersymmetric covariant derivative.

We write the supersymmetric Starobinsky Lagrangian in  the following form:
$L_S=L_1+\alpha L_2$, where $L_1$ is given by (\ref{linear}), and~\begin{equation}\label{streal}
L_2=\frac{3}{\kappa^2} \int d\Theta \mathcal{E} \mathcal{A}^3 \mathcal{R} \nabla_\theta \mathcal{R}.
\end{equation}

Integrating over the $\Theta$ variable, we~obtain the following:
\begin{eqnarray}\label{realsta}
L_S=\frac{3}{\kappa^2} N a^3 \left[-\frac{\dot{N}\dot{a}}{N^3a}+\frac{\ddot{a}}{N^2 a}+\frac{\dot{a}^2}{N^2 a^2}+\alpha \left(-\frac{\dot{N}\dot{a}}{N^3a}+\frac{\ddot{a}}{N^2 a}+\frac{\dot{a}^2}{N^2 a^2} \right)^2+i \frac{\lambda \dot{\psi}}{N}-\frac{i \dot{a}}{Na} \psi \lambda-i \frac{\psi \dot{\lambda}}{N}-i \frac{\lambda \dot{\lambda}}{N} \right. \nonumber \\
+\alpha \left(i \frac{\dot{\lambda} \ddot{\lambda}}{N^3}+\frac{9i \dot{N} {\dot{a}}^{2}}{N^4 a^2} \psi \lambda-\frac{8i \dot{a} \ddot{a}}{N^3 a^2} \psi \lambda-\frac{9i {\dot{a}}^{3}}{N^3 a^3} \psi \lambda+\frac{7i {\dot{a}}^{2}}{N^3 a^2} \lambda \dot{\lambda}+\frac{2i \dot{a}}{N^3 a} \lambda \ddot{\lambda}-\frac{i \dot{N} \dot{a}}{N^4a} \lambda \dot{\lambda}-\frac{i \ddot{a} \dot{a}}{N^3 a^2} \psi \lambda \right. \nonumber \\
+4i \frac{\dot{N} \dot{a}}{N^4 a} \psi \dot{\lambda}-\frac{4i \ddot{a}}{N^3 a} \psi \dot{\lambda}-\frac{i {\dot{a}}^{2}}{N^3 a^2} \psi \dot{\lambda}+4 \psi \lambda \frac{\dot{\psi} \dot{\lambda}}{N^2}- \frac{i \dot{a}}{N^3 a} \dot{\psi} \dot{\lambda}-\frac{i \ddot{a}}{N^3 a}\psi \dot{\lambda}+\frac{4i {\dot{a}}^{2}}{N^3 a^2} \lambda \dot{\psi}-\frac{i \ddot{a}}{N^3 a} \lambda \dot{\lambda} \nonumber \\
\left. \left. +\frac{i \dot{a}}{N^3 a} \psi \ddot{\lambda}+\frac{i {\dot{a}}^{2}}{N^3 a^2} \psi \dot{\psi}-2i \frac{\dot{N} \dot{a}}{N^4 a} \lambda \dot{\psi}+\frac{2i \ddot{a}}{N^3a} \lambda \dot{\psi} \right) \right].
\end{eqnarray}

The scale factor satisfies the expected fourth-order equation of motion, now with fermionic contributions. On~the other hand, the~term $\dot{\lambda} \ddot{\lambda}$ yields a third-order equation of motion. There is a tripling of fermionic degrees of freedom, compared to the ordinary case, as~we require  not only the initial value of $\lambda$, but~also those of its first two time derivatives. (The classical equations of motion are merely formal. Although~Grassmann algebras can be represented ``classically'' by matrices, the~proper treatment of a theory with fermions requires the algebra of quantum operators.)

The purely bosonic fourth-order equation of motion can be written in terms of the Hubble parameter, and~the third-order equation as follows:
\begin{equation}
0=\dddot{H}+6H \ddot{H}+\frac{9}{2}{\dot{H}}^{2}+9{H}^{2} \dot{H}+M^2 \left( \frac{3}{2}{H}^{2}+\dot{H}\right). \label{third}\\
\end{equation}

\begin{figure}
	\includegraphics[width=0.7\textwidth]{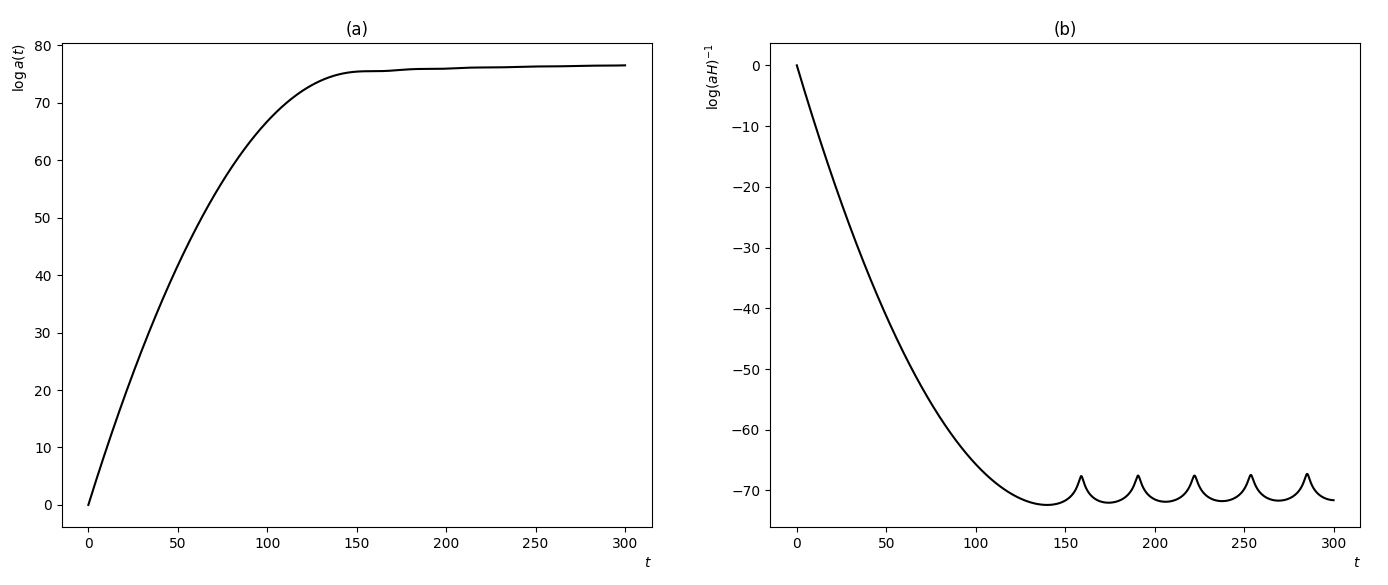}
	\caption{Numerical solution to (\ref{third}) with initial values \protect $a=1$, \protect $H=5M$, \protect $\dot{H}=-\frac{1}{6} M^2$, \protect $\ddot{H}=0$ and \protect $M=0.2$. (\textbf{a}) Logarithm of \protect $a(t)$; (\textbf{b}) comoving Hubble length (scale horizon).}
	\label{fig:Mesh2}
\end{figure}

Figure~\ref{fig:Mesh2} shows a numerical solution to (\ref{third}) displaying inflation during the large curvature regime. Initial values are chosen to satisfy the slow-roll conditions, and we set $\kappa=1$. The~scale factor increases from the order of $10^0$ up to $10^{32}$, which corresponds to, roughly, $73$ e-folds. From~that point, we obtain an oscillating amplitude superimposed on an overall expansion but~without inflation (this corresponds to the flattened part of the curve in Figure~\ref{fig:Mesh2}a). At~the final time displayed, we have in total 76 e-folds, approximately.

\unskip

\subsection{Scalar-Tensor~Formulation}\label{scalart}
As for the bosonic action (\ref{starobinsky}), a~Hamiltonian formulation can be obtained in different ways, and~the simplest is the one of type BF; Equation~(\ref{firststar}).
In our case, we can rewrite the Lagrangian density of (\ref{streal}) in~terms of $\mathcal{R}$ and another odd parity superfield $\Phi=\eta+\Theta \phi$ (with $\eta \eta=0$), as~\begin{equation}\label{last}
\mathcal{L}_2^\Phi=\mathcal E\mathcal A^3 (2\mathcal R-\Phi)\nabla_\theta \Phi+\mathcal E\Phi \mathcal R \nabla_\theta \mathcal A^3.
\end{equation}

We can check the equivalence at the superfield level. Using (\ref{realeq2}), we write the superfield equation of motion as $2 \mathcal{A}^3 \nabla_\theta (\mathcal{R}-\Phi)+(\nabla_\theta \mathcal{A}^3) (\mathcal{R}-\Phi)=0$. Therefore, $\Phi=\mathcal{R}$, and, substituting this back into (\ref{last}), we recover the integrand in the r.h.s. of (\ref{streal}). Note that the last term on the r.h.s of (\ref{last}) does not contribute when we replace $\Phi$ by $\mathcal{R}$, as~$\mathcal{R}$ is~an odd parity superfield.

Therefore, the~total equivalent scalar--tensor action has a Lagrangian density as follows:
\begin{equation}
\mathcal{L}_S^\Phi=\frac{3}{\kappa^2} \mathcal{E} \mathcal{A}^3 \left[ \mathcal{R}+\alpha \left( (2\mathcal R-\Phi)\nabla_\theta \Phi+3 \mathcal{A}^{-1} \Phi \mathcal{R} \nabla_\theta \mathcal A\right) \right],
\end{equation}

\noindent which~yields, after integrating by parts, the following:
\begin{eqnarray}\label{lagrangianreal}
L^\phi=\frac{3Na^3}{\kappa^2} \left[-\frac{\dot{a}^{2}}{a^2 N^2} \left(1+2\alpha \phi \right)-2\alpha \frac{\dot{a}\dot{\phi}}{aN^2}-\alpha {\phi}^{2} +2i \frac{\dot{a}}{a N} \psi \lambda -i\frac{\lambda \dot{\lambda}}{N}+\alpha \left(-3i \frac{\dot{a}\phi}{aN} \psi \lambda+7i \frac{\dot{a}\dot{\eta} \lambda}{a N^2}-2i \frac{\phi \psi \dot{\lambda}}{N} \right. \right. \nonumber \\
\left. \left. +2i \frac{\dot{\eta} \dot{\lambda}}{N^2}-2i \frac{\dot{a} \psi \dot{\eta}}{aN^2}+2i \frac{\dot{\phi}\psi \lambda}{N}+i \frac{\phi \lambda \dot{\lambda}}{N}-i \frac{\eta \dot{\eta}}{N}+3i \phi\eta \lambda+9i \frac{\dot{a}^{2}  \eta \lambda}{a^2 N^2}-6\frac{\psi \eta \lambda \dot{\lambda}}{N}+6i \frac{\dot{a} \eta \dot{\lambda}}{N^2}-3i \frac{\dot{a}^{2}\psi \eta}{a^2 N^2} \right) \right].
\end{eqnarray}

This Lagrangian contains two boson scalars, $a$ and $\phi$, and~two (real) fermion scalars, $\lambda$ and $\eta$. 
One can verify that the equations of motion for $\phi$ and $\eta$ are solved by $\eta\doteq -2 \frac{\dot{a}}{a} \lambda-\dot{\lambda}$ and $\phi \doteq  \left(\frac{1}{6}R+2i \lambda \dot{\lambda} \right)$, as~expected from the superfield~solution.

The absence of auxiliary fields greatly simplifies the expressions and allows us to construct a model whose bosonic part reproduces exactly the flat FLRW model of Starobinsky. Since the curvature superfield (\ref{superreal}) has odd parity, it probably does not correspond to the four-dimensional curvature superfield, and it could be related to its covariant derivatives. In~Section~\ref{cf}, we perform the Hamiltonian~formulation.

\section{N=2 Locally Supersymmetric~Action}\label{n2}
Now, we increase the dimension of the superspace by promoting $\Theta$ to a complex Grassmann variable ($\bar{\Theta}\equiv \Theta^*$, and~$\Theta \bar{\Theta}+\bar{\Theta} \Theta=0$, $\Theta \Theta=0=\bar{\Theta} \bar{\Theta}$). The basic results for this superspace are summarized in Appendix \ref{sec:A2}. Superfields have four components; one of them is usually an auxiliary field. For~convenience, we define the real scale factor superfield (the imaginary unit appearing in (\ref{a}) is a matter of convention, and we can dispense with it by redefining the $\lambda$s, e.g., $i(\Theta \bar{\lambda}+\bar{\Theta} \lambda) \to \Theta \lambda-\bar{\Theta} \bar{\lambda}$) as follows:
\begin{equation}\label{a}
\mathcal{A}(t,\Theta,\bar{\Theta})\equiv a(t) [1+i\Theta \bar{\lambda}(t)+i\bar{\Theta} \lambda(t)-\Theta \bar{\Theta} (s(t)-\lambda(t) \bar{\lambda}(t))],
\end{equation}
where $\bar{\lambda}=\lambda^*$. We write the scale factor superfield this way, which differs from the usual $\Theta$-expansion (\ref{auxsuper}) such that the lowest component of the curvature superfield $\mathcal{R}$ (defined below) is simply given by $s$. The~supersymmetry transformation of the supermultiplet $(a, \lambda, s)$ is non-linear and~can be obtained from (\ref{complextrans}).

The $k=0$ curvature superfield ({one can include positive spatial curvature adding $\sqrt{k} \mathcal{A}^{-1}$ to (\ref{supercurvature})}) is defined as the following:
\begin{equation}\label{supercurvature}
\mathcal{R}=\frac{1}{2}\mathcal{A}^{-1} [\nabla_{\bar{\theta}},\nabla_\theta]\mathcal{A}+\mathcal{A}^{-2} \nabla_{\bar{\theta}} \mathcal{A} \nabla_\theta \mathcal{A},
\end{equation}

Thus, we~have the following:
\begin{eqnarray}\label{supercurvature2}
\mathcal{R}=s+\Theta \left(\frac{2 \dot{a}}{Na}\bar{\lambda}+\frac{\dot{\bar{\lambda}}}{N}-\bar{\psi} \frac{\dot{a}}{Na}-i \psi \bar{\psi} \bar{\lambda}-i s \bar{\psi}-2 i s \bar{\lambda}\right)-\bar{\Theta} \left(\frac{2 \dot{a}}{Na}\lambda+\frac{\dot{\lambda}}{N}-\psi \frac{\dot{a}}{Na}+i \psi \bar{\psi} \lambda+i \psi s+2i s \lambda \right) \nonumber \\
+\Theta \bar{\Theta} \left(-\frac{\dot{N} \dot{a}}{N^3a}+\frac{\ddot{a}}{N^2a}+\frac{\dot{a}^2}{N^2a^2}+2{s}^{2}-i \frac{\dot{\psi} \bar{\lambda}+\dot{\bar{\psi}} \lambda}{N}-\frac{6i \dot{a}}{Na} (\psi \bar{\lambda}+\bar{\psi} \lambda)-2i \frac{\psi \dot{\bar{\lambda}}+\bar{\psi} \dot{\lambda}}{N}-2\psi \bar{\psi} s-4\psi \bar{\psi} \lambda \bar{\lambda} \right. \nonumber \\
\left. -2s (\psi \bar{\lambda}-\bar{\psi} \lambda)-2i \frac{\lambda \dot{\bar{\lambda}}+\bar{\lambda} \dot{\lambda}}{N}-8\lambda \bar{\lambda} s\right).
\end{eqnarray}

To better appreciate the content of large supersymmetric expressions, for~simplicity, we use the flat superspace gauge 
$N=1, \psi=0$ (this gauge corresponds to global supersymmetry; to check invariance under local supersymmetry (\ref{supergravity})--(\ref{complextrans}), actions must be written without gauge fixing, as~we do in Appendix \ref{app2}). 
In~the following, we indicate this gauge by equality with a dot, $\doteq$. Thus, the~Lagrangian for pure FLRW-supersymmetric cosmology is   the following:
\begin{equation}\label{L1}
L_1=\frac{3}{\kappa^2} \int d\Theta d\bar{\Theta} \mathcal{EA}^3 \mathcal{R} \doteq \frac{3a^3}{\kappa^2} \left(\frac{\ddot{a}}{a}+\frac{\dot{a}^2}{a^2}-s^{2}+i (\lambda \dot{\bar{\lambda}}+\bar{\lambda} \dot{\lambda})+s \lambda \bar{\lambda}\right),
\end{equation}
with the scalar density $\mathcal{E}$ given in (\ref{density}). Note that the superfield form of this action does not coincide with the one of previous works, e.g.,~\citep{garcia}.

In analogy with the $f(R)$ actions, we can write a superfield Lagrangian proportional to some function $F(\mathcal{R})$ (see~\cite{ketov2011} for the corresponding analysis in four-dimensional supergravity). From~(\ref{supercurvature2}) we have  the following:

\begin{eqnarray}\label{fdr}
L_F=\frac{3}{\kappa^2} \int d\Theta d\bar{\Theta} \mathcal{EA}^3 F(\mathcal{R}) 	\doteq \frac{3 a^3}{\kappa^2} \left[ F'(s) \left(\frac{\ddot{a}}{a}+\frac{\dot{a}^2}{a^2}+2{s}^{2}+i (\lambda \dot{\bar{\lambda}}+\bar{\lambda} \dot{\lambda}) +4s\lambda \bar{\lambda}\right)-3 F(s) (s+\lambda \bar{\lambda}) \right. \nonumber \\
\left. -F''(s) \left(\dot{\lambda} \dot{\bar{\lambda}}+4 {s}^{2} \lambda \bar{\lambda}+2 \frac{\dot{a}}{a}  (\lambda \dot{\bar{\lambda}}-\bar{\lambda} \dot{\lambda})+4 \frac{\dot{a}^2}{a^2} \lambda \bar{\lambda}+2 i  s (\lambda \dot{\bar{\lambda}}+\bar{\lambda} \dot{\lambda})\right) \right].
\end{eqnarray}

The equation of motion of the auxiliary field $s$ is of at least the second order in $s$, and~its solution leads to actions whose bosonic sectors are nonpolynomial in $R$. We have considered several examples of this type of actions, and~it seems that they contain ghosts, or~fail to produce large-curvature inflation. Hence, we will not discuss here this sort of~action.

The experience with the N=1 case of Section~\ref{n1} suggests to define an action depending on the covariant derivatives of the curvature superfield. Taking into account that $\nabla_\theta \mathcal{R}$ is an odd parity complex superfield, and~the Lagrangian density must be real and of even parity, the~natural candidate is $\nabla_\theta \mathcal{R} \nabla_{\bar{\theta}} \mathcal{R}$, which is the superfield kinetic term that is used to introduce scalar supersymmetric matter~\cite{tkach2}. As~in the case with real fermions, we write the following:
\begin{equation}\label{complexst}
L=L_1+\alpha L_2,
\end{equation}
with $L_1$ given in (\ref{L1}) and the following:
\begin{equation}\label{higherder}
L_2=\frac{3}{\kappa^2} \int d\Theta d\bar{\Theta}\ \mathcal{EA}^3 \nabla_{\bar \theta} \mathcal{R} \nabla_{\theta} \mathcal{R}.
\end{equation}

Performing the integration over $\Theta$-variables and summing, we~obtain the following:

\begin{eqnarray}\label{L2}
L \doteq \frac{3}{\kappa^2} a^3 \left[\frac{\ddot{a}}{a}+\frac{\dot{a}^2}{a^2}+\alpha \left(\frac{\ddot{a}}{a}+\frac{\dot{a}^2}{a^2}\right)^2+\alpha \dot{s}^{2}-s^{2}+4\alpha \left(\frac{\ddot{a}}{a}+\frac{\dot{a}^2}{a^2}\right) s^{2} +4\alpha {s}^{4}+i (\lambda \dot{\bar{\lambda}}+\bar{\lambda} \dot{\lambda})+s \lambda \bar{\lambda}+\alpha \left(-7 s \dot{\lambda} \dot{\bar{\lambda}} \right. \right. \nonumber \\
+2 s (\lambda \ddot{\bar{\lambda}}-\bar{\lambda} \ddot{\lambda})-i (\dot{\lambda} \ddot{\bar{\lambda}}+\dot{\bar{\lambda}} \ddot{\lambda})+i \frac{\ddot{a}}{a} (\lambda \dot{\bar{\lambda}}+\bar{\lambda} \dot{\lambda})-2 i \frac{\dot{a}}{a} (\lambda \ddot{\bar{\lambda}}+\bar{\lambda} \ddot{\lambda})-6 \frac{\dot{a}}{a} s (\lambda \dot{\bar{\lambda}}-\bar{\lambda} \dot{\lambda})-7i \frac{\dot{a}^2}{a^2} (\lambda \dot{\bar{\lambda}}+\bar{\lambda} \dot{\lambda}) \nonumber \\
\left. \left. +\lambda \bar{\lambda} \dot{\lambda} \dot{\bar{\lambda}}+\dot{s} (\lambda \dot{\bar{\lambda}}-\bar{\lambda} \dot{\lambda})-24 \frac{\dot{a}^2}{a^2} s \lambda \bar{\lambda}-20 {s}^{3} \lambda \bar{\lambda}+4 \frac{\ddot{a}}{a} s \lambda \bar{\lambda}+4 \frac{\dot{a}}{a} \dot{s} \lambda \bar{\lambda}-12i {s}^{2} (\lambda \dot{\bar{\lambda}}+\bar{\lambda} \dot{\lambda})\right)   \right].
\end{eqnarray}

Thus, with~the choice (\ref{higherder}), the total Lagrangian (\ref{complexst}) contains the FLRW model of Starobinsky as~well as the terms associated to the scalar field $s$. Further, the~kinetic term $\dot{s}^2$ in (\ref{L2}) tells us that the former auxiliary $s$ is promoted to a dynamical field. It is directly coupled to the curvature by the term $\frac{2}{3}\alpha Rs^2$ and has potential $V(s)=s^2-4 \alpha s^4$. The~equations of motion are of the fourth order for $a$ and second order for $s$. In~the fermionic sector, we find $\dot{\lambda} \ddot{\bar{\lambda}}+\dot{\bar{\lambda}} \ddot{\lambda}$ yielding third-order equations of~motion.

The scalar potential is unbounded for large $s$ but has a local minimum around $s=0$. However, since the effective quadratic mass is $M^2-\frac{2}{3}R$ (of the canonically normalized field $\tilde{s}=\frac{\sqrt{6}}{\kappa M} s$), we require not only $s$, but~also $R$ to be sufficiently small in order to obtain stable dynamics. On~the other hand, if~we set the initial conditions to $s=0=\dot{s}$, we can obtain inflation as in Section~\ref{n1}. However, this is not a stable solution in the sense that nonvanishing initial values of $s$ or $\dot{s}$, however small they are, eventually cause the field amplitude of $s$ to blow up while $a$ goes to~zero.

Up to now, we have partially succeeded in the construction of a supersymmetric model of Starobinsky: the bosonic sector of (\ref{L2}) already contains $R+\frac{\alpha}{6} R^2$, and we have no restriction on the value of $R$. However, we have a scalar field with a generally unstable potential. This is certainly not an unlikely situation; inflationary models derived from supergravity, and~more fundamental theories, e.g.,~string theory, contain several scalar fields. It is required that all of them but one sit in a stable vacuum state during inflation. This is, of~course, not the generic situation; it may happen that some of the additional fields develop instabilities, pushing the overall dynamics away from the inflationary solution. This is the case with theories of extra dimensions; after compactification, one needs a mechanism to stabilize the module fields~\cite{mcallister}.

To avoid fine tuning of the initial conditions, we add to the Lagrangian a term $\mathcal{R}^3$ to cancel the coupling $Rs^2$, which cancels also the negative sign fourth-power potential in \mbox{(\ref{L2})}. Thus, we propose the following superfield Lagrangian:
\begin{equation}\label{lagrangian}
\mathcal{L}_S=\frac{3}{\kappa^2} \mathcal{EA}^3 \left[\mathcal{R}+\alpha \left(\nabla_{\bar{\theta}} \mathcal{R} \nabla_{\theta} \mathcal{R}-\frac{4}{3} \mathcal{R}^3\right) \right],
\end{equation}
which yields the following:
\begin{eqnarray}\label{lagrangianboson}
L_S \doteq \frac{3a^3}{\kappa^2} \left[\frac{\ddot{a}}{a}+\frac{\dot{a}^2}{a^2}+\alpha \left(\frac{\ddot{a}}{a}+\frac{\dot{a}^2}{a^2}\right)^2+\alpha \dot{s}^{2}-s^{2}+i (\lambda \dot{\bar{\lambda}}+\bar{\lambda} \dot{\lambda})+s \lambda \bar{\lambda}+\alpha \left(\dot{s} (\lambda \dot{\bar{\lambda}}-\bar{\lambda} \dot{\lambda}) \right. \right. \nonumber \\
+s \dot{\lambda} \dot{\bar{\lambda}}+\lambda \bar{\lambda} \dot{\lambda} \dot{\bar{\lambda}}-i (\dot{\lambda} \ddot{\bar{\lambda}}+\dot{\bar{\lambda}} \ddot{\lambda})-2 i \frac{\dot{a}}{a} (\lambda \ddot{\bar{\lambda}}+\bar{\lambda} \ddot{\lambda})+i \frac{\ddot{a}}{a} (\lambda \dot{\bar{\lambda}}+\bar{\lambda} \dot{\lambda})+2 s (\lambda \ddot{\bar{\lambda}}-\bar{\lambda} \ddot{\lambda}) \nonumber \\
\left. \left. -7i \frac{\dot{a}^2}{a^2} (\lambda \dot{\bar{\lambda}}+\bar{\lambda} \dot{\lambda})+10 \frac{\dot{a}}{a} s (\lambda \dot{\bar{\lambda}}-\bar{\lambda} \dot{\lambda})+4 \frac{\dot{a}}{a} \dot{s} \lambda \bar{\lambda}+4 \frac{\ddot{a}}{a} s \lambda \bar{\lambda}+8 \frac{\dot{a}^2}{a^2} s \lambda \bar{\lambda} \right) \right].
\end{eqnarray}

See Appendix \ref{app2}, Equation~(\ref{lagrangianbosoncom}), for~the complete~expression. 

\begin{figure}
	\includegraphics[width=0.7\textwidth]{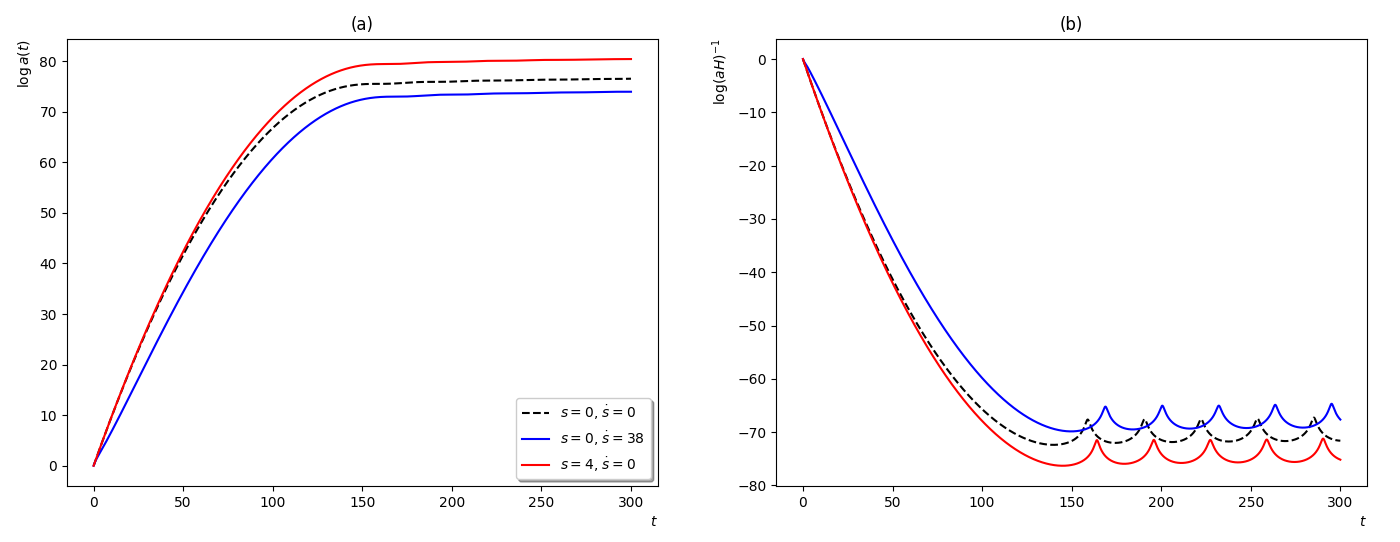}
	\caption{Numerical solutions to Equations~(\ref{scalareq}). The~initial conditions for the scale factor are the same as in Figure~\ref{fig:Mesh2}. Here, we have an additional scalar field. (\textbf{a}) Scale factor. (\textbf{b}) Comoving Hubble length for pure kinetic initial energy (blue color) and pure potential initial energy (red color) of the field \protect $s$. The~dotted line is the same as in the pure Starobinky dynamics of Figure~\ref{fig:Mesh2}.}\label{fig2}
\end{figure}

Thus, we obtain, in~the bosonic part, Starobinsky plus a massive scalar field. ({With $M$ constrained to be about $10^{-6} M_P$, $s$ is a rather heavy field. Considered in a field-theory context, heavy fields are more benevolent with the picture provided by the standard model of cosmology than lighter fields~\cite{mcallister}.}). 
The equations of motion of the bosonic sector~read as follows:
\begin{subequations}\label{scalareq}
	\begin{eqnarray}
	&&0=\frac{3}{2}{H}^{2}M^2+9{H}^{2} \dot{H}+\dot{H}M^2+\frac{9}{2} \dot{H}^2+6H \ddot{H}+\dddot{H}+\frac{\kappa^2 M^2}{4}  (\dot{\tilde{s}}^{2}-M^2 \tilde{s}^2) \label{eq6}, \\
	&&0=\ddot{\tilde{s}}+3H\dot{\tilde{s}}+M^2 \tilde{s}, \label{esq}
	\end{eqnarray}
\end{subequations}
where $\tilde{s}$ is the canonically normalized field, $\tilde{s}=\sqrt{6\alpha/\kappa^2}s$. Numerical solutions to Equations~(\ref{scalareq}) are shown in Figure~\ref{fig2} for different initial values of $s$ and $\dot{s}$. As~inflation takes place, the~field is driven to the minimum of the potential. Kinetic energy $M^2 \dot{\tilde{s}}^2$ is quickly dissipated by the friction term $H\dot{\tilde{s}}$ on the left-hand side of (\ref{esq}). In~contrast, this same friction term sustains a high value of the field for longer (acting as a cosmological constant), so we get more inflation (measured in e-folds) when most of the initial energy of $s$ is of the potential~type.

\unskip

\subsection*{Scalar--Tensor~Formulation}
Here, we write an equivalent Lagrangian following Sections~\ref{bef} and~\ref{scalart} such that it contains at most first-order time derivatives. We write in terms of $\mathcal{R}$ and a real scalar superfield $\Phi=\phi+i\Theta \bar{\eta}+i \bar{\Theta} \eta+\Theta \bar{\Theta} G$ the following:
\begin{equation}\label{firstorderreal}
\mathcal{L}_S^\Phi=\frac{3}{\kappa^2} \mathcal{EA}^3 \left[\mathcal{R}+\alpha \left(\nabla_{\bar{\theta}} \Phi \nabla_{\theta} \mathcal{R}-\nabla_{\theta} \Phi \nabla_{\bar{\theta}} \mathcal{R}-\nabla_{\bar{\theta}} \Phi \nabla_{\theta} \Phi-\frac{4}{3} \mathcal{R}^3\right) \right].
\end{equation}

The equivalence can be verified using the superfield equation of motion for $\Phi$. From~\mbox{(\ref{realeq})}, we obtain $\nabla_{\bar{\theta}} \mathcal{A}^3 \nabla_\theta (\mathcal{R}-\Phi)-\nabla_\theta \mathcal{A}^3 \nabla_{\bar{\theta}} (\mathcal{R}-\Phi)-\mathcal{A}^3 [\nabla_{\theta},\nabla_{\bar{\theta}}] (\mathcal{R}-\Phi)=0$. Therefore, the~solution is given by $\Phi=\mathcal{R}+c$, where $c$ is a constant. Replacing $\Phi$ by $\mathcal{R}+c$ into (\ref{firstorderreal}) returns (\ref{higherder}). Performing the fermionic integration in the action yields, for the $\Phi$-dependent part, the following:
\begin{eqnarray}\label{kinetic}
L^{\Phi} \doteq \frac{3a^3}{\kappa^2} \left[2 \left(\frac{\ddot{a}}{a}+\frac{\dot{a}^2}{a^2}\right) G-{G}^{2}+4G {s}^{2}+2 \dot{\phi} \dot{s}- {\dot{\phi}}^{2}+i  (\eta \dot{\bar{\eta}}+\bar{\eta} \dot{\eta})+2 (\dot{\lambda} \dot{\bar{\eta}}-\dot{\bar{\lambda}} \dot{\eta})+3 s \eta \bar{\eta} \right. \nonumber \\
+3i  \lambda \bar{\lambda} (\dot{\lambda} \bar{\eta}+\dot{\bar{\lambda}} \eta)-G i   (\lambda \dot{\bar{\lambda}}+\bar{\lambda} \dot{\lambda})-3 i  s (\dot{\lambda} \bar{\eta}+\dot{\bar{\lambda}} \eta)+4 \frac{\dot{a}}{a} (\lambda \dot{\bar{\eta}}-\bar{\lambda} \dot{\eta})+3 \frac{\dot{a}}{a} (\dot{\lambda} \bar{\eta}-\dot{\bar{\lambda}} \eta) \nonumber \\
+3 \dot{\phi} (\lambda \dot{\bar{\lambda}}-\bar{\lambda} \dot{\lambda})+4i  s (\lambda \dot{\bar{\eta}}+\bar{\lambda} \dot{\eta})+3G (\lambda \bar{\eta}-\bar{\lambda} \eta)+3 \frac{\dot{a}^2}{a^2} (\lambda \bar{\eta}-\bar{\lambda} \eta)+3i  \dot{s} (\lambda \bar{\eta}+\bar{\lambda} \eta) \nonumber \\
\left. +12 \frac{\dot{a}}{a} \dot{\phi} \lambda \bar{\lambda}-3 \frac{\ddot{a}}{a} (\lambda \bar{\eta}-\bar{\lambda} \eta)-3i  \dot{\phi} (\lambda \bar{\eta}+\bar{\lambda} \eta)+3\lambda \bar{\lambda} \eta \bar{\eta}-4Gs \lambda \bar{\lambda}\right].
\end{eqnarray}

As expected from the superfield expression (\ref{firstorderreal}) and (\ref{complextrans}), this Lagrangian depends only on $\dot\phi$; hence it can be eliminated. On~the other hand, it can be seen that $G$ plays the role of the scalaron, and~we eliminate the coupling $Gs^2$ by the shift $G \to G'=G-2s^2$.
Thus, renaming $G'=G$, the~Lagrangian reads as follows:
\begin{eqnarray}\label{lagrangiancomplex}
L_S^G\doteq \frac{3a^3}{\kappa^2} \left[ \left(\frac{\ddot{a}}{a}+\frac{\dot{a}^{2}}{a^2}\right) (1+2\alpha G)-\alpha G^{2} +\alpha {\dot{s}}^{2}-{s}^{2}+i (\lambda \dot{\bar{\lambda}}+\bar{\lambda} \dot{\lambda})+s \lambda \bar{\lambda}+\alpha \left(8 s \dot{\lambda} \dot{\bar{\lambda}} \right. \right.  \nonumber \\
+ i (\eta \dot{\bar{\eta}}+\bar{\eta} \dot{\eta})+2 (\dot{\lambda} \dot{\bar{\eta}}-\dot{\bar{\lambda}} \dot{\eta})-3i  s (\dot{\lambda} \bar{\eta}+\dot{\bar{\lambda}} \eta)+4 i  s (\lambda \dot{\bar{\eta}}+\bar{\lambda} \dot{\eta})+16 \frac{\dot{a}}{a} s (\lambda \dot{\bar{\lambda}}-\bar{\lambda} \dot{\lambda}) \nonumber \\
+10 i  {s}^{2} (\lambda \dot{\bar{\lambda}}+\bar{\lambda} \dot{\lambda})+3 \dot{s} (\lambda \dot{\bar{\lambda}}-\bar{\lambda} \dot{\lambda})+7 \frac{\dot{a}}{a} (\lambda \dot{\bar{\eta}}-\bar{\lambda} \dot{\eta})+12 \frac{\dot{a}}{a} \dot{s} \lambda \bar{\lambda}+6 \frac{\dot{a}}{a} (\dot{\lambda} \bar{\eta}-\dot{\bar{\lambda}} \eta) \nonumber \\
-\frac{9}{2} \lambda \bar{\lambda} \dot{\lambda} \dot{\bar{\lambda}}+9 \frac{\dot{a}^{2}}{a^2} (\lambda \bar{\eta}-\bar{\lambda} \eta)+\frac{15}{2} i  \lambda \bar{\lambda} (\dot{\lambda} \bar{\eta}+\dot{\bar{\lambda}} \eta)-\frac{3}{2}\lambda \bar{\lambda} \eta \bar{\eta}+6 {s}^{2} (\lambda \bar{\eta}-\bar{\lambda} \eta) \nonumber \\
\left. \left. -Gi (\lambda \dot{\bar{\lambda}}+\bar{\lambda} \dot{\lambda})+3s \eta \bar{\eta}+3G (\lambda \bar{\eta}-\bar{\lambda} \eta)-4Gs \lambda \bar{\lambda}+12{s}^{3} \lambda \bar{\lambda} +32 \frac{\dot{a}^{2}}{a^2} s \lambda \bar{\lambda}\right) \right].
\end{eqnarray}

In Appendix \ref{app2}, Equation~(\ref{finalcomplex}), we show the full~expression.

\section{Canonical~Formulation}\label{cf}
For the main actions worked out in this article, namely, (\ref{realsta}) and (\ref{lagrangianboson}), we have classically equivalent actions, (\ref{lagrangianreal}) and (\ref{lagrangiancomplex}), which still contain terms that are quadratic in the fermionic velocities: $\dot{\eta} \dot{\lambda}$ in (\ref{lagrangianreal}) and $\dot{\lambda} \dot{\bar{\eta}}-\dot{\bar{\lambda}} \dot{\eta}$ in (\ref{lagrangiancomplex}). They are, however, sufficiently good to perform the usual Hamiltonian formulation. Typical fermionic Lagrangians contain, at most, linear terms in the velocities, which ultimately lead to second-class constraints of the form $\pi_\lambda=\partial L/\partial \dot{\lambda}=\bar{\lambda}$ \cite{ramirez2016}. For~the present case, quadratic velocity terms allow us to solve for all the (physical) velocities, bosonic and fermionic, in~terms of coordinates and momenta. There still arise some constraints, but~all of them are first class and~form a closed algebra under the Poisson bracket extended to fermionic variables (as defined in~\cite{henneaux}). 

For the case with real fermions, with~Lagrangian (\ref{realsta}), there are two primary (first-class) constraints $p_N=0$, $\pi_\psi=0$. The~rest of the conjugate momenta can be solved for the velocities. Thus, for~the Hamiltonian $H_S^\phi=\dot{N}p_N+\dot{\psi} \pi_\psi+\dot{a} p_a+\dot{\phi} p_\phi+\dot{\lambda} \pi_\lambda+\dot{\eta} \pi_\eta-L$, we obtain the following:
\begin{equation}\label{totalr}
H_S^\phi=N H_0+\frac{1}{2} \Psi S,
\end{equation}
where $\Psi=2N \psi$ (see Appendix A), where the Hamiltonian and supersymmetric constraints~are the following:

\begin{subequations}
	\fontsize{9}{9}\selectfont
	\begin{eqnarray}
	&&H_0=-\frac{\kappa^2}{6\alpha} \frac{p_{a} p_\phi}{a^2}+\frac{\kappa^2}{12 \alpha^2} (1+2\alpha \phi) \frac{p_\phi^{2}}{a^3}+\frac{3\alpha}{\kappa^2} {a}^{3} {\phi}^{2}+i \left( \frac{\kappa^2}{\alpha} \frac{p_\phi^{2}}{a^3}+\frac{3 {a}^{3}}{2\kappa^2} (1-7\alpha \phi)\right) \eta \lambda \nonumber \\
	&&\ \ \ \ \ \ \ \ \ \ \ +\frac{7 \kappa^2}{12 \alpha} \frac{p_\phi}{a^3} \lambda \pi_{\lambda}+\frac{1-\alpha \phi}{2\alpha} \lambda \pi_{\eta}+\frac{\kappa^2}{2\alpha} \frac{p_\phi}{a^3} \eta \pi_{\eta}+\frac{i \kappa^2}{6\alpha} \frac{\pi_{\lambda} \pi_{\eta}}{a^3}-\frac{\eta \pi_{\lambda}}{2},\label{hreal}\\
	&&S=i \left(a p_{a}-\frac{1-\alpha \phi}{2\alpha } p_\phi \right) \lambda-i \left(\frac{\kappa^2}{4\alpha} \frac{p_\phi^{2}}{a^3}+\frac{3\alpha}{\kappa^2} {a}^{3} \phi\right) \eta+\frac{\kappa^2}{6\alpha} \frac{p_\phi}{a^3} \pi_{\lambda}+\phi \pi_{\eta}. \label{sreal}
	\end{eqnarray}
\end{subequations}

Note that $S$ is imaginary and $H$ is real. The~algebra of constraints, all of which are first class, closes under Poisson brackets. In~particular, the~usual relation  $\{S,S\}=\frac{i}{2} H_0$~holds.

For the case of complex fermions, the~full Lagrangian corresponding to (\ref{lagrangiancomplex}) is given in the Appendix \ref{app2}. With~this Lagrangian, the bosonic momenta are defined in the usual way, although~we use a slightly different definition for fermions (mostly to keep a consistent notation, and the over-bar is still equivalent to complex conjugation): $\pi_\lambda=\partial{L}/\partial \dot{\bar{\lambda}}$, $\pi_{\bar{\lambda}}=-\partial L/\partial \dot{\lambda}$, and~so on. The~only primary constraints come from the momenta associated to the gauge fields: $p_N=0$, $\pi_\psi=0$, $\pi_{\bar{\psi}}=0$. Solving for the rest of velocities and computing $H=\dot{N}p_N-\dot{\psi} p_{\bar{\psi}}+\dot{\bar{\psi}} \pi_\psi+\dot{a} p_a+\dot{G} p_G+\dot{s} p_s-\dot{\lambda} \pi_{\bar{\lambda}}+\dot{\bar{\lambda}} \pi_\lambda-\dot{\eta} \pi_{\bar{\eta}}+\dot{\bar{\eta}} \pi_\eta-L$, we obtain the following:
\begin{equation}
H=N H_0+\frac{1}{2} (\Psi \bar{S}-\bar{\Psi} S),
\end{equation}
where

\begin{subequations}
\begin{eqnarray}
H_0=\frac{\kappa^2}{12 \alpha^2} (1+2\alpha G) \frac{p_G^{2}}{a^3}-\frac{\kappa^2}{6\alpha} \frac{p_a p_G}{a^2}+\frac{\kappa^2}{12\alpha} \frac{p_s^{2}}{a^3}+\frac{3{a}^{3}}{\kappa^2} \left( \alpha{G}^{2}+{s}^{2}\right)-\frac{18\alpha}{\kappa^2} {a}^{3} \lambda \bar{\lambda} \eta \bar{\eta}-\frac{3 \kappa^2}{4\alpha} \frac{p_G}{a^3} \left(p_s+2 p_G s\right) \lambda \bar{\lambda} \nonumber \\
+\frac{\kappa^2}{4\alpha a^3} \left(p_s+4s p_G \right) (\lambda \pi_{\bar{\eta}}-\bar{\lambda} \pi_\eta)-\frac{is}{2} (\eta \pi_{\bar{\eta}}+\bar{\eta} \pi_\eta)-\frac{7\kappa^2}{12}\frac{p_G}{a^3} (\lambda \pi_{\bar{\lambda}}-\bar{\lambda} \pi_\lambda)-\frac{3i}{4} (p_s+5p_G s) (\lambda \bar{\eta}+\bar{\lambda} \eta) \nonumber\\
+\frac{3 a^3}{2 \kappa^2} \left(1-7\alpha G-6 \alpha{s}^{2} \right)  (\lambda \bar{\eta}-\bar{\lambda} \eta)-\frac{\kappa^2}{2\alpha} \frac{p_G}{a^3} (\eta \pi_{\bar{\eta}}-\bar{\eta} \pi_\eta)+2is (\lambda \pi_{\bar{\lambda}}+\bar{\lambda} \pi_\lambda)+\frac{3}{\kappa^2} {a}^{3}s \left(3-4\alpha {s}^{2}\right) \lambda \bar{\lambda} \nonumber \\
+\frac{\kappa^2}{\alpha} \frac{p_G^{2}}{a^3} (\lambda \bar{\eta}-\bar{\lambda} \eta)+\frac{\kappa^2}{6\alpha a^3} (\pi_\lambda \pi_{\bar{\eta}}-\pi_{\bar{\lambda}} \pi_\eta)-\frac{i}{2} \left(G+6 {s}^{2}\right) (\lambda \pi_{\bar{\eta}}+\bar{\lambda} \pi_\eta)-\frac{2\kappa^2}{3\alpha} \frac{s}{a^3}\pi_\eta \pi_{\bar{\eta}}-\frac{6\alpha}{\kappa^2} {a}^{3} s \eta \bar{\eta} \nonumber \\
+\frac{i}{2} (\eta \pi_{\bar{\lambda}}+\bar{\eta} \pi_\lambda)-\frac{15i}{4} \lambda \bar{\lambda} (\eta \pi_{\bar{\eta}}+\bar{\eta} \pi_\eta)+\frac{i}{2\alpha} (\lambda \pi_{\bar{\eta}}+\bar{\lambda} \pi_\eta),\label{hcomplex}\\
S=\left(i a p_a- \frac{3}{\kappa^2} {a}^{3} s- \frac{\kappa^2}{4\alpha} \frac{p_G p_s}{a^3}+\frac{i}{2}G p_G-\frac{ip_G}{2\alpha}(1-6\alpha s^2)+\frac{12\alpha}{\kappa^2} {a}^{3} s (G+{s}^{2}) \right) \lambda+3\left(\frac{5}{4}i p_G +\frac{6\alpha}{\kappa^2} {a}^{3} s\right) \lambda \bar{\lambda} \eta \nonumber\\
+\frac{\kappa^2}{6\alpha} \frac{p_s}{a^3} \pi_\eta-2i \lambda \bar{\lambda} \pi_\lambda+\left(i s-\frac{\kappa^2}{6\alpha} \frac{p_G}{a^3}\right) \pi_\lambda+\left(\frac{i}{2} \left(p_s-3 p_G s\right)+\frac{\kappa^2}{4\alpha} \frac{p_G^{2}}{a^3}+\frac{3\alpha}{\kappa^2} {a}^{3} (G+2{s}^{2})\right) \eta+\frac{9\alpha}{2\kappa^2} {a}^{3} \lambda \eta \bar{\eta} \nonumber \\
-\frac{3i}{2} (\lambda \bar{\eta}+\bar{\lambda} \eta) \pi_\eta -i (G+2 {s}^{2}) \pi_\eta. \label{scomplex}
\end{eqnarray}
\end{subequations}

$\bar{S}$ is the complex conjugate of (\ref{scomplex}).
As in the previous case, we have the Poisson bracket relation $\{S,\bar{S}\}=\frac{i}{2} H_0$.

Both Hamiltonians, (\ref{hreal}) and (\ref{hcomplex}), have, after~diagonalization, a~negative kinetic term that corresponds to the usual gravitational instability. There are also terms lineal in the fermionic momenta. However, fermions do not have classical counterparts and~require a quantum analysis,  which will be detailed in future work.
Quantization is accomplished by promoting classical variables to operators satisfying (anti-)commutation relations. The~wave function must satisfy the Wheeler--DeWitt equation $H_0\Psi=0$, as~well as the supersymmetric constraint equations $S\Psi=0=\bar{S}\Psi$ \cite{ramirez2016}. By~virtue of the anticommutation relation $\{S,\bar{S}\}=\frac{i}{2}H_0$, a~solution to $S\Psi=0=\bar{S}\Psi$ automatically satisfies $H_0\Psi=0$. Usually, the~supersymmetric constraint amounts to a partial differential equation of a lower order than the Wheeler--DeWitt equation, which represents an enormous simplification when looking for analytic solutions~\cite{moniz}. That is not quite the case here since there are terms that are quadratic in the momenta of the scalars in (\ref{sreal}) and (\ref{scomplex}), although~the equations are significantly~simpler.

\section{Conclusions}\label{concl}
We presented two supersymmetric extensions of the FLRW model of Starobinsky, with~real and complex fermions, using a superfield formalism for 1D supergravity. It was shown that the bosonic sectors of these actions allow for the large-curvature\linebreak inflationary~solution.

In the case of N=1 supersymmetry, Section~\ref{n1}, computations are very simple because the supermultiplets only contain two components and no auxiliary fields. Despite the small number of degrees of freedom, it is possible to construct a Lagrangian whose bosonic sector contains exactly $R+\frac{\alpha}{6}R^2$. It is a property of 1D supermultiplets that all the components can be physical; in more dimensions, auxiliary fields are, in~general, inevitable. Then, we considered an N=2 complex supermultiplet containing two scalar bosons and one complex scalar fermion. In~theories without higher derivatives, one of the bosons is an auxiliary field, but~in our case with higher-derivative terms, it becomes~dynamical.

In comparison to N=1, the~N=2 model required more maneuvering; the superfield kinetic term generates $R^2$, and~promotes the lowest component of the curvature superfield to a dynamical field. However, this new field comes equipped with a negative quartic potential, preventing the inflationary solution. We fixed this by adding a superpotential term of the form $F(\mathcal{R})$. With~the choice $F(\mathcal{R})=-8\mathcal{R}^3$, the~scalar potential of this extra field was significantly improved. The~final Lagrangian contains, besides~FLRW Starobinsky, a~minimally coupled massive scalar field. The~classical dynamics of the bosonic part of the action, obtained numerically, shows $R^2$-driven inflation, while the scalar field $s$ remains in a low-energy~state.

We wrote equivalent actions for the two models, by~including additional bosonic and fermionic fields. First, we obtained formal superfield expressions, which were verified and further developed at the component level. Let us recall here that these alternative formulations are at an intermediate level, in~the sense that they are not yet in the form of standard supersymmetric theories. Nonetheless, they are already suitable for the Hamiltonian formulation. The~Hamiltonian expressions contain the purely bosonic model of Starobinsky and, in~the N=2 case, contain an additional massive scalar~field.

From our discussion above, the~full tensor--scalar duals of our models would require two additional superfields to accommodate the three fermionic degrees of freedom. The~manifestly supersymmetric way to arrive at that formulation should involve the superfield generalization of the Weyl rescaling, and~redefinition of the scalaron field. However, as~we pointed out, in~our models, it is the highest component of the extra superfield that plays the role of the scalaron. Hence, the~proper superfield transformation should involve not $\Phi$, but~its covariant derivatives. An~adequate derivation of the transformation rules, including fermions, will be addressed in subsequent work. Nonetheless, we can anticipate that their bosonic sectors will be of the form (\ref{frwstar}). For~the N=2 model (\ref{hcomplex}), since the field $s$ is canonically normalized in the frame of Starobinsky, in~the Einstein frame, its contribution would appear in the form of a non-linear sigma model (cf. the two-field model in \citep{ketovn}).

Finally, the~actions proposed in this work belong to supersymmetric, also called ``pseudo'', classical mechanics because the dynamical variables are elements of Grassmann algebras. These actions find application as quantum theories, that is, quantum supersymmetric cosmology \citep{moniz}. The~quantization of these models, along the lines of Refs.~\citep{ramirez2016,garcia} and~its comparison with the pure bosonic case~\cite{hawkingluttrell,vazquez,ramirez2018}, will be the topic of upcoming work. Among~other interesting aspects to be investigated, is deriving the modified Friedmann equation, reflecting the effect of fermions by~means of a semi-classical approach \citep{escamilla}.

\appendix
\section{Local Supersymmetry in~Superspace}\label{app}
Superspace provides a geometrical realization of supersymmetry and supergravity. For~the models constructed in this work, we use the covariant formulation of one-dimensional supergravity provided by the new $\Theta$ variables formalism. We summarize here the basic expressions we used; for~more details, see~\cite{wessbagger}, and~especially~\cite{ramirez,garcia}.

\subsection{N=1 D=1~Superspace}
It has local coordinates $z^M=(t, \Theta)$. $t$ is the ordinary time coordinate, whereas $\Theta$ is a real Grassmann parameter, $\Theta^*=\Theta$, $\Theta \Theta=0$. Functions valued on this space or superfields can be expanded in a finite Taylor series in the $\Theta$ coordinate. A~real superfield is given by  the following:
\begin{equation}\label{realsuper}
\Phi(t,\Theta)=\phi(t)+i \Theta \eta(t).
\end{equation}
where $\eta$ is a real, odd parity, Grassmann~variable.

The vielbein superfields $\nabla_M^{\ A}(t,\Theta)$ provide a coordinate independent basis of 1-forms, $E^A=d^M\nabla_M^{\ A}$, $A=\tau, \theta$. For~N=1 1D, we have the following:
\begin{eqnarray}
&\nabla_{M}^{\ A}=\begin{bmatrix}
N-i\Theta \Psi && \frac{1}{2}\Psi \\
i\Theta && 1
\end{bmatrix}. \label{vielreal} 
\end{eqnarray}	

Covariant derivatives are defined with the inverse vielbein fields $\nabla_A^{\ M}$ as $\nabla_A \Phi=\nabla_A^{\ M} \partial_M \Phi$ (for N=1, 2 the following relations hold: $\nabla_M^{\ A}\nabla_A^{\ N}=\delta_M^{\ N}$, $\nabla_A^{\ M} \nabla_M^{\ B}=\delta_A^{\ B}$). They are given~by the following:
\begin{subequations}\label{realcov}
	\begin{eqnarray}
	&&\nabla_\tau \Phi=N^{-1} \dot{\phi}-i \psi \eta+i\Theta N^{-1} (\dot{\eta}+ \dot{\phi} \psi),\\
	&&\nabla_\theta \Phi=i\eta-\Theta (i N^{-1} \dot{\phi}+\psi \eta).
	\end{eqnarray}	
\end{subequations}
where $\psi \equiv \frac{1}{2N} \Psi$.

Under a local supersymmetry transformation parameterized by $\zeta(t)$, the~one-dimensional supergravity multiplet transforms as the following:
\begin{align}\label{realgauget}
\delta_\zeta N=-i \zeta \Psi, && \delta_\zeta \Psi=-2 \dot{\zeta}.
\end{align}
whereas, for~the real supermultiplet contained in (\ref{realsuper}), we have the following:
\begin{align}\label{realtrans}
\delta_\zeta \phi=-i \zeta \eta, &&\delta_\zeta \eta=\zeta (N^{-1} \dot{\phi}-i \psi \eta).
\end{align}

Supersymmetric invariant actions are of the form ({Berezin integration rules are $\int d\Theta (\phi+\Theta \eta)=\eta$. For~the complex case, $\int d\Theta d\bar{\Theta} (\phi+\Theta \eta-\bar{\Theta} \bar{\eta}+\Theta \bar{\Theta} G)=G$}).
\begin{equation}\label{action}
S=\int dt L\equiv \int dt d\Theta \mathcal{L}=\int dt d\Theta \mathcal{EJ}
\end{equation}
where $\mathcal{E}$ is a scalar density superfield~\cite{wessbagger}. Here, we use  the following, given the super-determinant of (\ref{vielreal}):
\begin{equation}\label{realdensity}
\mathcal E=N(1-i\Theta \psi).
\end{equation}

Finally, formal superfield equations of motion can be readily derived. For~an odd parity superfield $\Gamma$, such as the one used in Section~\ref{n1}, we have the following:
\begin{equation}\label{realeq2}
\nabla_\theta \frac{\partial \mathcal{J}}{\partial \nabla_\theta \Gamma}+\frac{\partial \mathcal{J}}{\partial \Gamma}=0
\end{equation}

Compared to the case of even parity, there is an extra minus sign in the first term of \mbox{(\ref{realeq2})}, which arises from $\nabla_\theta (\delta \Gamma \dots)=-\delta \Gamma \nabla_\theta \dots$.

\subsection{N=2 D=1~Superspace}\label{sec:A2}
This superspace has local coordinates $z^M=(t,\Theta,\bar \Theta)$, with~$\Theta \Theta=0=\bar{\Theta} \bar{\Theta}$, $\Theta \bar{\Theta}+\bar{\Theta} \Theta=0$. In~comparison to the previous one, we have promoted $\Theta$ to a complex Grassmann number (we use an over-bar to denote complex conjugation, which for products is defined with an order exchange, e.g.,~$(\lambda \gamma)^*=\bar{\gamma} \bar{\lambda}$) and defined $\Theta^*=\bar{\Theta}$. An~arbitrary superfield has a finite Taylor expansion as follows:
\begin{equation}\label{auxsuper}
\Phi(t,\Theta,\bar \Theta)=\phi(t)+i \Theta \bar{\eta}(t)+i\bar{\Theta} \eta (t)+\Theta \bar{\Theta} G(t),
\end{equation}
The vielbein reads as follows:
\begin{eqnarray}
\nabla_M^{\ A}=\begin{bmatrix}
N+i (\Theta \bar{\Psi}+\bar{\Theta} \Psi) && \frac{1}{2}\Psi && \frac{1}{2}\bar \Psi \  \\
-i\bar \Theta && 1 && 0 \\
i\Theta && 0 && -1
\end{bmatrix}, \label{viel}
\end{eqnarray}
Covariant derivatives are given by the following:
\begin{equation}\label{deriv1}
\nabla_\theta \Phi=i\bar{\eta}+\bar \Theta (i \dot{\phi}/N+\psi \bar{\eta}+\bar{\psi} \eta+G)+\Theta \bar{\Theta} (\dot{\bar{\eta}}/N-\bar{\psi} \dot{\phi}/N-i \psi \bar{\psi} \bar{\eta}+i \bar{\psi} G),
\end{equation}
with $\nabla_{\bar{\theta}} \Phi$ given by the complex conjugate of (\ref{deriv1}) (as before, we defined $\psi\equiv \frac{1}{2N}\Psi$). Under~a local supersymmetry transformation parameterized by $\zeta(t), \bar{\zeta}(t)$,
\begin{align}\label{supergravity}
\delta_\zeta N=-i(\zeta \bar \Psi+\bar \zeta \Psi), &&\delta_\zeta \Psi=-2\dot{\zeta},
\end{align}
whereas the following holds:
\begin{subequations}\label{complextrans}
	\begin{eqnarray}
	&&\delta_\zeta \phi=i(\bar{\eta} \zeta+\eta \bar{\zeta}),\\
	&&\delta_\zeta \eta=-\zeta \frac{\dot{\phi}}{N}-iG \zeta+i (\psi \bar{\eta}+\bar{\psi} \eta) \zeta,\\
	&&\delta_\zeta G=\frac{\zeta \dot{\bar{\eta}}-\bar{\zeta} \dot{\eta}}{N}-\frac{\dot{\phi}}{N} (\psi \bar{\zeta}-\bar{\psi} \zeta)-i G (\psi \bar{\zeta}+\bar{\psi} \zeta)+i \psi \bar{\psi} (\eta \bar{\zeta}+\bar{\eta} \zeta) \label{t3}.
	\end{eqnarray}
\end{subequations}
The superspace action is $S=\int dt L\equiv \int dt d\Theta d\bar{\Theta} \mathcal{L}=\int dt d\Theta d\bar{\Theta} \mathcal{EJ}$, with the following:
\begin{equation}\label{density}
\mathcal{E}=-N (1+i \Theta \bar{\psi}+i \bar{\Theta} \psi).
\end{equation}
An even parity superfield satisfies the equation of motion as follows:
\begin{equation}\label{realeq}
\nabla_{\theta} \frac{\partial \mathcal{J}}{\partial \nabla_{\theta} \Phi}+\nabla_{\bar{\theta}} \frac{\partial \mathcal{J}}{\partial \nabla_{\bar{\theta}} \Phi}-\frac{\partial \mathcal{J}}{\partial \Phi}=0.
\end{equation}


\section{Full~Lagrangians}\label{app2}
In this appendix, we give lengthy Lagrangians that are shown in a short~form. The full Lagrangian corresponding to  (\ref{lagrangianboson}) is the following:
\begin{eqnarray}\label{lagrangianbosoncom}
	L_S=\frac{3}{\kappa} N a^3 \left[\frac{\ddot{a}}{N^2a}-\frac{\dot{N} \dot{a}}{N^3a}+\frac{\dot{a}^2}{N^2a^2}+\alpha \left(\frac{\ddot{a}}{N^2 a}-\frac{\dot{N}\dot{a}}{N^3a}+\frac{\dot{a}^2}{N^2 a^2} \right)^2+\frac{\alpha {\dot{s}}^{2}}{N^2}-{s}^{2}+s \lambda \bar{\lambda}-i \frac{\dot{\psi} \bar{\lambda}+\dot{\bar{\psi}} \lambda}{N}+2\psi \bar{\psi} \lambda \bar{\lambda} \right. \nonumber \\
	-\frac{i\dot{a}}{Na} (\psi \bar{\lambda}+\bar{\psi} \lambda)-i \frac{\psi \dot{\bar{\lambda}}+\bar{\psi} \dot{\lambda}}{N}+i \frac{\lambda \dot{\bar{\lambda}}+\bar{\lambda} \dot{\lambda}}{N}+\alpha \left(s\frac{\dot{\lambda} \dot{\bar{\lambda}}}{N^2}-i \frac{\dot{\lambda} \ddot{\bar{\lambda}} +\dot{\bar{\lambda}} \ddot{\lambda}}{N^3}+2 s \frac{\lambda \ddot{\bar{\lambda}}-\bar{\lambda} \ddot{\lambda}}{N^2}+\frac{i \ddot{a}}{N^2a} \frac{\lambda \dot{\bar{\lambda}}+\bar{\lambda} \dot{\lambda}}{N} \right. \nonumber\\
		-\frac{2 i \dot{a}}{Na} \frac{\lambda \ddot{\bar{\lambda}}+\bar{\lambda} \ddot{\lambda}}{N^2} -\frac{7i \dot{a}^2}{N^2 a^2} \frac{\lambda \dot{\bar{\lambda}}+\bar{\lambda} \dot{\lambda}}{N}+\frac{10 s \dot{a}}{Na} \frac{\lambda \dot{\bar{\lambda}}-\bar{\lambda} \dot{\lambda}}{N}-\frac{11 s \dot{a}^{2}}{N^2a^2} (\psi \bar{\lambda}-\bar{\psi} \lambda)-\frac{s \dot{a}}{Na} \frac{\psi \dot{\bar{\lambda}}-\bar{\psi} \dot{\lambda}}{N}-\frac{6 s \ddot{a}}{N^2a} \psi \bar{\psi} \nonumber\\
			-\frac{4 i \dot{a}^{2}}{N^2a^2} \frac{\dot{\psi} \bar{\lambda}+\dot{\bar{\psi}} \lambda}{N}-9 i \frac{\dot{a}^{3}}{N^3 a^3} (\psi \bar{\lambda}+\bar{\psi} \lambda)-9i \frac{\dot{a} \ddot{a}}{N^3 a^2} (\psi \bar{\lambda}+\bar{\psi} \lambda)+i \frac{\dot{N} \dot{a}}{N^3a} \frac{\lambda \dot{\bar{\lambda}}+\bar{\lambda} \dot{\lambda}}{N}+2i \frac{\dot{N} \dot{a}}{N^3a} \frac{\dot{\psi} \bar{\lambda}+\dot{\bar{\psi}} \lambda}{N} \nonumber
	\end{eqnarray}
\begin{eqnarray}
	+\frac{\dot{s}}{N} \frac{\lambda \dot{\bar{\lambda}}-\bar{\lambda} \dot{\lambda}}{N}-\frac{2 i \ddot{a}}{N^2 a} \frac{\dot{\psi} \bar{\lambda}+\dot{\bar{\psi}} \lambda}{N}-i{s}^{2} \frac{\psi \dot{\bar{\lambda}}+\bar{\psi} \dot{\lambda}}{N}+\frac{6s\dot{N} \dot{a}}{N^3a} \psi \bar{\psi}+\frac{6 \dot{s} \dot{a}}{N^2 a} \psi \bar{\psi}+\frac{\lambda \ddot{\bar{\lambda}}-\bar{\lambda} \ddot{\lambda}}{N^2} \psi \bar{\psi} +\frac{i \dot{a}}{Na} \frac{\psi \ddot{\bar{\lambda}}+\bar{\psi} \ddot{\lambda}}{N^2} \nonumber \\
	-\frac{7 \dot{a} \dot{s}}{N^2a} (\psi \bar{\lambda}-\bar{\psi} \lambda)+9 is \frac{\psi \dot{\bar{\lambda}}+\bar{\psi} \dot{\lambda}}{N} \lambda \bar{\lambda}-2 \frac{\dot{\lambda} \dot{\bar{\lambda}}}{N^2} (\psi \bar{\lambda}-\bar{\psi} \lambda)-s\frac{\dot{\psi} \dot{\bar{\lambda}}-\dot{\bar{\psi}} \dot{\lambda}}{N^2}-2\frac{\dot{\psi} \dot{\bar{\psi}}}{N^2} \lambda \bar{\lambda}-\frac{22 \dot{a}}{N a} \psi \bar{\psi} \frac{\lambda \dot{\bar{\lambda}}-\bar{\lambda} \dot{\lambda}}{N} \nonumber \\
	- \frac{s\dot{N}}{N^2} \frac{\psi \dot{\bar{\lambda}}+\bar{\psi} \dot{\lambda}}{N}+\frac{4 \dot{s} \dot{a}}{N^2 a} \lambda \bar{\lambda}-\frac{53 \dot{a}^{2}}{N^2a^2} \psi \bar{\psi} \lambda \bar{\lambda}-5 \frac{\psi \dot{\bar{\psi}} \lambda \dot{\bar{\lambda}}}{N^2}-5 \frac{\bar{\psi} \dot{\psi} \bar{\lambda} \dot{\lambda}}{N^2}+\frac{\psi \dot{\bar{\psi}} \bar{\lambda} \dot{\lambda}}{N^2}+\frac{\bar{\psi} \dot{\psi} \lambda \dot{\bar{\lambda}}}{N^2}-\frac{3\dot{s}}{N} \frac{\psi \dot{\bar{\lambda}}-\bar{\psi} \dot{\lambda}}{N} \nonumber \\
	-3{s}^{3} \psi \bar{\psi}-\frac{s \ddot{a}}{N^2a} (\psi \bar{\lambda}-\bar{\psi} \lambda)-3i s \psi \bar{\psi} \frac{\lambda \dot{\bar{\lambda}}+\bar{\lambda} \dot{\lambda}}{N}+s \frac{\psi \ddot{\bar{\lambda}}-\bar{\psi} \ddot{\lambda}}{N^2}-4 \frac{\psi \dot{\psi} \bar{\lambda} \dot{\bar{\lambda}}}{N^2}-4 \frac{\bar{\psi} \dot{\bar{\psi}} \lambda \dot{\lambda}}{N^2}+\frac{3 \dot{a}}{Na} \frac{\psi \dot{\bar{\lambda}}-\bar{\psi} \dot{\lambda}}{N} \lambda \bar{\lambda} \nonumber \\
	+\frac{\dot{\lambda} \dot{\bar{\lambda}}}{N^2} \lambda \bar{\lambda}+\frac{s \dot{N} \dot{a}}{N^3 a} (\psi \bar{\lambda}-\bar{\psi} \lambda)-\frac{4 \dot{a} s}{Na} \frac{\dot{\psi} \bar{\lambda}-\dot{\bar{\psi}} \lambda}{N}-\frac{2 \dot{N} \dot{a}}{N^3 a} \psi \bar{\psi} \lambda \bar{\lambda}-\frac{13 \dot{a}}{N a} \frac{\psi \dot{\bar{\psi}}-\bar{\psi} \dot{\psi}}{N} \lambda \bar{\lambda}+\frac{4 i \dot{N} \dot{a}}{N^3a} \frac{\psi \dot{\bar{\lambda}}+\bar{\psi} \dot{\lambda}}{N} \nonumber \\
	-12  \psi \bar{\psi} \frac{\dot{\lambda} \dot{\bar{\lambda}}}{N^2}-\frac{i \dot{a}}{N a} \frac{\dot{\psi} \dot{\bar{\lambda}}+\dot{\bar{\psi}} \dot{\lambda}}{N^2}-\frac{5 i \ddot{a}}{N^2 a} \frac{\psi \dot{\bar{\lambda}}+\bar{\psi} \dot{\lambda}}{N}+\frac{3 s \dot{a}^{2}}{N^2a^2} \psi \bar{\psi}-4 \frac{s \dot{N} \dot{a}}{N^3a} \lambda \bar{\lambda}+\frac{4 s \ddot{a}}{N^2a} \lambda \bar{\lambda}+i s \frac{\psi \dot{\bar{\psi}}+\bar{\psi} \dot{\psi}}{N} \lambda \bar{\lambda} \nonumber \\
	+\frac{8 \dot{a}^{2} s}{N^2a^2} \lambda \bar{\lambda}+7 i \frac{s \dot{s}}{N} (\psi \bar{\lambda}+\bar{\psi} \lambda)+6is \psi \bar{\psi} \frac{\dot{\psi} \bar{\lambda}+\dot{\bar{\psi}} \lambda}{N}+2i s^2 \frac{\dot{\psi} \bar{\lambda}+\dot{\bar{\psi}} \lambda}{N}+6i \frac{{s}^{2} \dot{a}}{Na} (\psi \bar{\lambda}+\bar{\psi} \lambda)+9 {s}^{2} \psi \bar{\psi} \lambda \bar{\lambda} \nonumber \\
	-\frac{2\dot{a}}{N a} \frac{\dot{\psi} \bar{\lambda}-\dot{\bar{\psi}} \lambda}{N} \psi \bar{\psi}-\frac{\dot{N}}{N^2} \frac{\lambda \dot{\bar{\lambda}}-\bar{\lambda} \dot{\lambda}}{N} \psi \bar{\psi}+\frac{2 \ddot{a}}{N^2a} \psi \bar{\psi} \lambda \bar{\lambda}+\frac{\dot{\psi} \dot{\bar{\lambda}}-\dot{\bar{\psi}} \dot{\lambda}}{N^2} \lambda \bar{\lambda}+\frac{9 i \dot{N} \dot{a}^2}{N^4a^2} (\psi \bar{\lambda}+\bar{\psi} \lambda) \nonumber \\
	\left. \left. -\frac{i \dot{a}^2}{N^2 a^2} \frac{\psi \dot{\bar{\lambda}}+\bar{\psi} \dot{\lambda}}{N}-i {s}^{2} \frac{\psi \dot{\bar{\psi}}+\bar{\psi} \dot{\psi}}{N}-\frac{i \dot{a}^2}{N^2 a^2} \frac{\psi \dot{\bar{\psi}}+\bar{\psi} \dot{\psi}}{N}-2 \frac{\dot{N}a}{N^2} \frac{\lambda \dot{\bar{\lambda}}-\bar{\lambda} \dot{\lambda}}{N} \right) \right].
\end{eqnarray}
The full Lagrangian corresponding to (\ref{lagrangiancomplex}), once we integrate by parts, which is used to compute the Hamiltonian (\ref{hcomplex}), is the following:
\begin{eqnarray}\label{finalcomplex}
	L_S^G=\frac{3Na^3}{\kappa^2} \left[\frac{-\dot{a}^{2}}{N^2 a^2} (1+2\alpha G)-2 \frac{\alpha \dot{a} \dot{G}}{N^2 a}-\alpha {G}^{2}+\frac{\alpha \dot{s}^{2}}{N^2}-{s}^{2}+i \frac{\lambda \dot{\bar{\lambda}}+\bar{\lambda} \dot{\lambda}}{N}+2\psi \bar{\psi} \lambda \bar{\lambda}+s \lambda \bar{\lambda}+2i \frac{\dot{a}}{N a} (\psi \bar{\lambda}+\bar{\psi} \lambda) \right. \nonumber \\
	+\alpha \left(2 \frac{\dot{\lambda} \dot{\bar{\eta}}-\dot{\bar{\lambda}} \dot{\eta}}{N^2}+3\frac{\dot{\lambda} \dot{\bar{\lambda}}}{N^2} (\psi \bar{\lambda}-\bar{\psi} \lambda)-3i s \frac{\dot{\lambda} \bar{\eta}+\dot{\bar{\lambda}} \eta}{N}+\frac{9\dot{a}^{2}}{N^2 a^2} (\lambda \bar{\eta}-\bar{\lambda} \eta)+10 i {s}^{2} \frac{\lambda \dot{\bar{\lambda}}+\bar{\lambda} \dot{\lambda}}{N}+\frac{3\dot{s}}{N} \frac{\lambda \dot{\bar{\lambda}}-\bar{\lambda} \dot{\lambda}}{N} \right. \nonumber \\
	+8 s\frac{\dot{\lambda} \dot{\bar{\lambda}}}{N^2}+6 \frac{i s\dot{a}}{Na} (\psi \bar{\eta}+\bar{\psi} \eta)+\frac{7i s\dot{s}}{N} (\psi \bar{\lambda}+\bar{\psi} \lambda)-iG \frac{\lambda \dot{\bar{\lambda}}+\bar{\lambda} \dot{\lambda}}{N}+2i s \frac{\psi \dot{\bar{\eta}}+\bar{\psi} \dot{\eta}}{N}+12{s}^{3} \lambda \bar{\lambda}-\frac{2 \dot{s}}{N} \frac{\psi \dot{\bar{\lambda}}-\bar{\psi} \dot{\lambda}}{N} \nonumber \\
	+4i {s}^{2} \frac{\psi \dot{\bar{\lambda}}+\bar{\psi} \dot{\lambda}}{N}+4 i s \frac{\lambda \dot{\bar{\eta}}+\bar{\lambda} \dot{\eta}}{N}-\frac{\dot{a}^{2} \psi \bar{\psi}}{2 N^2 a^2} \lambda \bar{\lambda}-\frac{2\dot{a}}{N a} \frac{\psi \dot{\bar{\eta}}-\bar{\psi} \dot{\eta}}{N}-\frac{8s \dot{a}}{N a} \frac{\psi \dot{\bar{\lambda}}-\bar{\psi} \dot{\lambda}}{N}+18i \frac{{s}^{2} \dot{a}}{N a} (\psi \bar{\lambda}+\bar{\psi} \lambda) \nonumber \\
	-2G \psi \bar{\psi} \lambda \bar{\lambda}-4Gs \psi \bar{\psi}-\frac{9i \dot{a}}{2Na} (\psi \bar{\eta}+\bar{\psi} \eta) \lambda \bar{\lambda}+3 (\psi \bar{\lambda}-\bar{\psi} \lambda) \eta \bar{\eta}+3G (\lambda \bar{\eta}-\bar{\lambda} \eta)-\frac{9}{2}\lambda \bar{\lambda} \frac{ \dot{\lambda} \dot{\bar{\lambda}}}{N^2}-4Gs \lambda \bar{\lambda} \nonumber \\
	+\frac{3i \psi \bar{\lambda} \dot{\bar{\lambda}} \eta}{N}-3i\frac{\psi \lambda \dot{\bar{\lambda}} \bar{\eta}}{N}+\frac{3i \bar{\psi} \bar{\lambda} \dot{\lambda} \eta}{N}- \frac{3i \bar{\psi} \lambda \dot{\lambda} \bar{\eta}}{N}+\frac{32s\dot{a}^{2}}{a^2 N^2} \lambda \bar{\lambda}-\frac{6i \psi \bar{\lambda} \dot{\lambda} \bar{\eta}}{N}+\frac{6i \eta \bar{\psi} \lambda \dot{\bar{\lambda}}}{N}-\frac{3i G \dot{a}}{Na} (\psi \bar{\lambda}+\bar{\psi} \lambda) \nonumber \\
	+3i \frac{\psi \dot{\bar{\eta}}+\bar{\psi} \dot{\eta}}{N} \lambda \bar{\lambda}+2i \frac{\dot{G}}{N} (\psi \bar{\lambda}+\bar{\psi} \lambda)-\frac{3\dot{a}^{2}}{N^2a^2} (\psi \bar{\eta}-\bar{\psi} \eta)+\frac{7\dot{a}}{Na} \frac{\lambda \dot{\bar{\eta}}-\bar{\lambda} \dot{\eta}}{N}+3s \eta \bar{\eta}-2i G \frac{\psi \dot{\bar{\lambda}}+\bar{\psi} \dot{\lambda}}{N} \nonumber \\
	-\frac{21}{2} i s \frac{\psi \dot{\bar{\lambda}}+\bar{\psi} \dot{\lambda}}{N} \lambda \bar{\lambda}+6{s}^{3} (\psi \bar{\lambda}-\bar{\psi} \lambda)-\frac{9}{2} s^{2} \psi \bar{\psi} \lambda \bar{\lambda}+\frac{6\dot{a}}{Na} \frac{\dot{\lambda} \bar{\eta}-\dot{\bar{\lambda}} \eta}{N}+\frac{8s\dot{a}^{2}}{N^2 a^2} \psi \bar{\psi}+6 {s}^{2} (\lambda \bar{\eta}-\bar{\lambda} \eta) \nonumber \\
	-Gs (\psi \bar{\lambda}-\bar{\psi} \lambda)+i \frac{\eta \dot{\bar{\eta}}+\bar{\eta} \dot{\eta}}{N}-\frac{\dot{a} \psi \bar{\psi}}{N a} \frac{\lambda \dot{\bar{\lambda}}-\bar{\lambda} \dot{\lambda}}{N}+\frac{16 s \dot{a}}{Na} \frac{\lambda \dot{\bar{\lambda}}-\bar{\lambda} \dot{\lambda}}{N}-\frac{16 s\dot{a}^{2}}{N^2a^2} (\psi \bar{\lambda}-\bar{\psi} \lambda)+\frac{4\dot{a} \dot{s}}{N^2a} \psi \bar{\psi} \nonumber \\
	-\frac{27}{2}s (\psi \bar{\eta}-\bar{\psi} \eta) \lambda \bar{\lambda}+\frac{12 \dot{s}\dot{a}}{N^2a} \lambda \bar{\lambda}+i s \psi \bar{\psi} \frac{\lambda \dot{\bar{\lambda}}+\bar{\lambda} \dot{\lambda}}{N}+3 {s}^{2} (\psi \bar{\eta}-\bar{\psi} \eta)-\frac{3}{2}\lambda \bar{\lambda} \eta \bar{\eta}-\frac{3\dot{a}}{2Na} \frac{\psi \dot{\bar{\lambda}}-\bar{\psi} \dot{\lambda}}{N} \lambda \bar{\lambda} \nonumber \\
	\left. \left. -2 \psi \bar{\psi} \frac{\dot{\lambda} \dot{\bar{\lambda}}}{N^2}+\frac{15i}{2} \frac{\dot{\lambda} \bar{\eta}+\dot{\bar{\lambda}} \eta}{N} \lambda \bar{\lambda}+2i \psi \bar{\psi} \frac{\lambda \dot{\bar{\eta}}+\bar{\lambda} \dot{\eta}}{N}-\frac{7\dot{a}\dot{s}}{N^2a} (\psi \bar{\lambda}-\bar{\psi} \lambda)+6s \psi \bar{\psi} (\lambda \bar{\eta}-\bar{\lambda} \eta) \right)\right].
\end{eqnarray}


\vskip 1truecm
\centerline{\bf Acknowledgements}
N.E. Mart\'{\i}nez-P\'erez thanks CONACyT for the studies grant during this work. C. R. thanks O. Obreg\'on and H. Garc\'{\i}a Compe\'an for the useful~discussions.

\end{document}